\RequirePackage[l2tabu, orthodox]{nag} 

\documentclass[11pt,fleqn]{article}

\usepackage[margin=1in]{geometry}
\usepackage[toc,page]{appendix} 

\usepackage[utf8]{inputenc}
\usepackage[T1]{fontenc}
\usepackage[english]{babel}

\usepackage{setspace}
\usepackage{fontenc} 
\usepackage{verbatim}

\newread\tmp

\usepackage{amsmath,amsthm,amssymb,scrextend,mathtools}
\usepackage{ragged2e}
\usepackage{etoolbox}
\usepackage{scalerel}
\usepackage[all,cmtip]{xy}
\usepackage{relsize}
\usepackage{colonequals} 
\usepackage{physics} 
\usepackage{mathrsfs}
\usepackage{xfrac} 
\usepackage{cases} 

\usepackage{tikz} 
\usepackage{tikz-cd} 
\usepackage{fancybox} 
\usepackage{pgfplots} 
\pgfplotsset{compat=1.18}
\usepackage{graphicx}
\usepackage{makecell} 

\usepackage{caption} 
\usepackage{subcaption} 
\usepackage{multicol} 
\usepackage{enumerate} 
\usepackage{multirow} 
\usepackage{booktabs}
\usepackage{array}
\usepackage{tabularx} 
\usepackage{float} 
\usepackage{wrapfig} 
\usepackage{xltabular}
\usepackage{makecell}

\newcolumntype{L}[1]{>{\raggedright\arraybackslash}p{#1}}
\newcolumntype{C}[1]{>{\centering\arraybackslash}p{#1}}
\newcolumntype{R}[1]{>{\raggedleft\arraybackslash}p{#1}}

\usepackage{enumitem}
\usepackage{setspace}
\usepackage{color}

\definecolor{green}{HTML}{4F783F}
\definecolor{lightgreen}{HTML}{f0f5f0}
\definecolor{orange}{HTML}{C9582C}
\definecolor{purple}{HTML}{A774AA}
\definecolor{darkpurple}{HTML}{4E0250}
\definecolor{yellow}{HTML}{F5BB00}
\definecolor{blue}{HTML}{1C7C9C}
\definecolor{lightblue}{HTML}{E6EDF2}

\definecolor{codebg}{HTML}{F7F7F7}
\definecolor{coderule}{HTML}{DDDDDD}

\usepackage{empheq}
\usepackage{svg} 
\usepackage[many]{tcolorbox} 
\usepackage{xparse} 

\usepackage[color=yellow,linecolor=black]{todonotes}

\usepackage{ulem} 

\usepackage{empheq} 

\usepackage{algorithm}
\usepackage[noend]{algpseudocode}

\makeatletter
\def\BState{\State\hskip-\ALG@thistlm}
\makeatother

\usepackage{listings} 
\usepackage{minted} 
\usepackage{xstring} 

\numberwithin{equation}{section} 
\numberwithin{figure}{section}   

\usepackage[square,comma,numbers,sort]{natbib}
\usepackage[colorlinks=true, pdfborder={0 0 0}]{hyperref}
\hypersetup{urlcolor=blue, citecolor=red}
\usepackage{url}

\newtheorem*{theorem*}{Theorem}

\renewcommand{\maketitle}{
    \begin{center}
        {\LARGE \bfseries \@title \par}
        \vspace{0.5em}
        {\large \@author \par}
        \vspace{0.5em}
        {\small \@date \par}
    \end{center}
    \vspace{1em}
}

\renewenvironment{abstract}{
    \begin{center}
        \textbf{Abstract}
    \end{center}
    \begin{quote}
}{
    \end{quote}
}

\usepackage{authblk} 
\usepackage{setspace}

\date{}

\makeatletter
\renewcommand{\maketitle}{
    \begin{center}
        {\LARGE \bfseries \@title \par}
        \vspace{1em}
        {\large \@author \par}
        \vspace{0.5em}
        {\small \@date \par}
    \end{center}
    \vspace{1.5em}
}
\makeatother

\usepackage{authblk}

\date{}

\usepackage{placeins} 

\begin{document}


\title{Infection, Immunity, and Population Dynamics in Midwestern Bat Hibernacula: A Boolean Network Model of White-Nose Syndrome}

\author[1]{Calum Heldt}
\affil[1]{\texttt{cheldt@unomaha.edu}, University of Nebraska at Omaha}

\author[2]{Nicholas Linares}
\affil[2]{\texttt{nlinares@unomaha.edu}, University of Nebraska at Omaha}

\author[3]{Mihaela Teodora Velcsov}
\affil[3]{\texttt{dvelcsov@unomaha.edu},
University of Nebraska at Omaha (Corresponding author)}

\maketitle


\setcounter{secnumdepth}{4} 
\setcounter{tocdepth}{4} 


\begin{abstract}
Since 2006, white-nose syndrome (WNS) has rapidly
spread across the continental United States and
devastated bat populations. Northern long-eared bats,
tricolored bats, and little brown bats have suffered
population declines exceeding 90\% in affected areas.
We develop a stochastic Boolean network model to
investigate the effects of WNS on tricolored bat
populations in hibernacula in the Midwestern United
States. The model incorporates transmission dynamics,
host resistance, and environmental fungal persistence.
Using sensitivity analyses, invasion scenarios, and
parameter phase diagrams, we examine short-term
infection dynamics and long-term population outcomes.
Our results indicate that transmission parameters
strongly influence infection dynamics, while acquired
resistance and environmental fungal decay play important
roles in long-term population survival. Initial infection
levels have relatively little influence on long-term
outcomes, and population declines may become apparent
only after several years.
\end{abstract}

\section{Introduction}\label{sec:intro}

White-nose syndrome (WNS) is caused by the fungus
\textit{Pseudogymnoascus destructans}
(\textit{P. destructans}), which thrives in the cool,
moist environments of bat hibernacula and poses a
serious threat to bats during hibernation
\cite{verant_WNS_physiology}.
The tricolored bat (\textit{Perimyotis subflavus})
is one of the three species particularly vulnerable
to WNS infection, with population declines exceeding
90\% in affected areas and some populations
completely extirpated \cite{cheng_scope_2021}.
\textit{P. destructans} infects exposed skin,
particularly the wings and muzzle, during torpor,
when bats are particularly vulnerable.
Infection can cause bats to arouse from hibernation
and search for food or water.
However, these arousals substantially deplete fat
reserves during a season when resources are scarce,
potentially leading to death from dehydration,
starvation, or exposure \cite{loeb_changes_2022}.

Environmental factors such as land use, mean air temperature,
and elevation are strongly correlated with the spread of WNS
\cite{flory_environmental_2012}. In particular,
\textit{P. destructans} grows preferentially in colder
environments with relative humidity above 70\%
\cite{loeb_changes_2022}. However, geographic location
alone does not capture the biological and environmental
interactions that may influence a population's susceptibility
to WNS. We investigate these interactions using a mathematical
model focused on tricolored bat hibernacula in the
Midwestern United States.

Previous studies have emphasized the importance of both host
responses and environmental conditions in the dynamics of
white-nose syndrome. In particular, environmental contamination
and pathogen persistence can contribute to transmission within
hibernacula \cite{hoyt_environmental_2020,langwig_invasion_2015}.
Capturing such interactions is a central challenge in ecological
modeling, where individual-based approaches can represent
differences among individuals and their responses to changing
conditions \cite{grimm2005pattern,grimm2006odm}.
Mathematical models of WNS have also investigated disease
transmission and environmental influences. For example, a susceptible--exposed--infected (SEI)
model has been developed to investigate WNS dynamics
while incorporating environmental reservoirs of
\textit{P. destructans} \cite{lilley_landscape_2018}.
\textit{Boolean network} (BN) approaches have also been used
to study ecological dynamics through simplified representations
of interacting components \cite{gaucherel_understand_2017}.
Motivated by these approaches, we develop a stochastic BN
that represents individual bats through inhabitant nodes
and incorporates environmental factors through environmental
nodes. Biologically motivated rules and transition probabilities
govern changes in bat states, while an environmental reservoir
accounts for fungal contamination within hibernacula.
Our framework provides a complementary approach for examining
how transmission, host resistance, and environmental fungal
persistence influence short-term disease dynamics and
long-term population outcomes.

The goal of this study is to better understand how
biological, environmental, and epidemiological factors
influence the survival, resistance, or extirpation of
bats in hibernacula affected by WNS. We use data from
Midwestern hibernacula, grouping the sites into North
and South Midwest regions based on latitude, to inform
parameter estimation. We investigate WNS dynamics
under different model conditions through sensitivity
analysis, invasion scenarios, and parameter phase
diagrams to identify factors that may contribute to
different population outcomes.

\noindent The following research questions motivate our investigation:
\begin{enumerate}
    \item How do transmission parameters, acquired resistance,
    and environmental fungal decay influence WNS dynamics
    and bat population outcomes?

    \item Which model parameters have the greatest influence
    on short-term disease dynamics and long-term population
    survival, and how does their importance differ between
    Midwestern regions?

    \item How do initial infection levels and combinations
    of model parameters affect long-term population
    trajectories, including delayed population declines?
\end{enumerate}

The paper is organized as follows.
Section~\ref{sec:Boolean_model} introduces the
stochastic BN model, including its nodes,
parameters, and update rules.
Section~\ref{sec:methods} describes the parameterization
of the model and the computational methods used for
parameter estimation and simulation.
Section~\ref{sec:analysis} presents the methods used to
analyze model behavior, including infection scenarios
and sensitivity analyses, together with the corresponding
simulation results.
Section~\ref{sec:discussion} discusses their
interpretation and implications.
Finally, Section~\ref{sec:conclusions} summarizes the
main conclusions and outlines directions for future work.

\section{Boolean Network Model}\label{sec:Boolean_model}

Boolean networks were originally introduced as discrete dynamical models for genetic regulation \cite{Kauffman}, where node states take binary values $0$ and $1$ representing simplified activity levels such as ``ON/OFF'' or ``active/inactive''. Early extensions, including probabilistic formulations, addressed uncertainty in biological regulation and led to probabilistic Boolean network models \cite{Shmulevich:02, Shmulevich:04, shmulevich_dougherty_pbn_book}. More generally, Boolean networks provide a natural framework for binary-state dynamical processes on complex networks, including both biological regulation and epidemics. In ecological settings, such binary-state dynamics naturally arise in host--pathogen systems, where individuals or populations transition between susceptible and infected states, as in other wildlife disease systems \cite{hoyt_ecology_2021, frick2010emerging, lorch2011pathology, lorch_distribution_2013}.

Recent developments further establish Boolean networks as predictive and data-driven models for gene regulatory systems and complex network dynamics \cite{kadelka2024_bn_meta, murrugarra2024_bn_control, Pusnik2022_bn_inference_review}. In systems biology, Boolean network models have been extensively developed for gene regulatory systems, where long-term attractors are interpreted as stable phenotypes, and modern approaches emphasize control and intervention strategies in complex networks \cite{Albert:03, Saadatpour:13, rozum2023_bn_predictive}. From a structural and dynamical perspective, Boolean networks also highlight the role of topology in determining global behavior, including redundancy, node influence, and entropy-based reduction. These properties have been studied through concepts such as determinative power of nodes and network reduction while preserving essential dynamics \cite{helikar2013comprehensive, pentzien2018_determinative_power, wacker2020_topology_determinative, pelz2021_entropy_reduction}.

Finally, in ecological theory, similar ideas appear in models of ecosystem dynamics, resilience, and regime shifts, where discrete or simplified representations of complex systems are used to capture large-scale transitions \cite{grimm2005pattern, gaucherel_understand_2017}. In this context, Boolean network formulations have been applied to ecological and social--ecological systems, including regime shift modeling in ecosystems \cite{gaucherel_understand_2017}, and can be extended to epidemic processes in ecological networks \cite{hoyt_ecology_2021, frick2010emerging, lorch2011pathology}, which we further apply to white-nose syndrome in bat populations.

However, even a Boolean simplification of reality can pose challenges in assessing the dynamics of the network due to the exponential dependence of the state space on the number of nodes. More precisely, a network with $N$ nodes generates $2^N$ possible states of the network, i.e., all possible combinations of $N$ elements with values of either 0 or 1. In the present model, we reduce the computational complexity by assuming that bats belonging to the same subcategory (as described in Section~\ref{sec:nodes}) have identical update rules. Under this assumption, nodes belonging to the same subcategory are interchangeable with respect to their update rules, although their individual trajectories may differ because state transitions are stochastic. This leads to more flexibility in the choice of parameters that may influence the behavior of the network, such as environmental factors. In mathematical terms, we consider a fully connected Boolean network consisting of a set of nodes $\{ x_1, x_2, \dots , x_N\} $ and a list of corresponding Boolean functions or update rules $(f_1, f_2, \dots , f_N)$. For simplicity, we assume that every bat can interact with every other bat, resulting in a fully connected network. The Boolean function $f_i(x_1, x_2, \dots , x_N)$ is assigned to node $x_i, i = 1, 2, \dots, N$. Accordingly, each update function satisfies $f_i : \{0,1\}^N \rightarrow \{0, 1\}$. We assume that $f_i = f_j$ for the nodes $x_i, x_j$ contained in the same subcategory of nodes. Since the network evolves in time, we can denote $x_i(t)$ to be the value of node $x_i$ at time $t$. Then to calculate the value of $x_i$ at time $t+1$, we apply its corresponding rule $f_i$ as a function of the state of the network at time $t$. That is, $x_i(t+1) = f_i\big(x_1(t), x_2(t), \dots , x_N(t)\big)$. We initialize the network at $t=0$, and stop it when $t$ reaches some pre-determined time $T>0$. This equation describes the classical deterministic Boolean network framework; in our model, state transitions additionally involve stochastic trials governed by transition probabilities. Moreover, each time step represents one day.

Our model is inspired by individual-based ecological modeling approaches such as
\cite{grimm2005pattern,grimm2006odm,railsback2019agent}, while adopting the Boolean network-based paradigm that has previously been used to study ecological dynamics and ecosystem development as in
\cite{gaucherel_understand_2017}.
The individual inhabitants of a hibernaculum are hence represented as nodes of a Boolean network whose state evolves according to biologically motivated logical rules. Environmental variables are incorporated as external nodes that influence state transitions, and stochasticity is introduced through Bernoulli trials governing state transitions, using probabilities suitable for ecological and epidemiological processes. The resulting framework preserves the logic of Boolean network modeling while capturing the stochastic interactions characteristic of wildlife disease systems.

Next, we describe the particular characteristics of the proposed Boolean network model for the bat hibernaculum ecosystem, including the nodes, the necessary parameters, and the update rules.

\subsection{Network Nodes and States}\label{sec:nodes}

To encapsulate all relevant dynamics within a bat hibernaculum, we define two distinct categories of nodes: inhabitant and environmental nodes. The inhabitants form the main Boolean network, while the environmental category provides external Boolean inputs to the evolving network. This categorical separation of ecological interactions comes from Gaucherel et al. \cite{gaucherel_understand_2017}.

In the proposed model, the inhabitant category represents all living and deceased bats sharing the hibernaculum. The population of tricolored bats -- on which this research is focused -- is divided into five behavioral subcategories: hibernating (Hi), non-hibernating and immune (Im), non-hibernating and infected (In), non-hibernating and neither immune nor infected (Ot), and deceased (De). Each subcategory of \textbf{inhabitant} nodes may contain multiple nodes, whose states at a given time indicate the individuals belonging to that subcategory. An exhaustive list of nodes is described in Table \ref{tab:nodes}. The inhabitant subcategories represent mutually disjoint sets of inhabitants.

In contrast, each subcategory of the \textbf{environment} is represented by a single node. The environment category contains relevant environmental factors known to influence bat hibernation and resource availability. These nodes affect the inhabitant network and are not influenced by the progression of the bat population in the model. The factors considered are temperature (Te), humidity (Hu), and resources (Re) \cite{loeb_changes_2022}. Temperature, humidity, and elevation have all been extensively studied in relation to WNS \cite{loeb_changes_2022}. We do not include elevation as a node because (1) small elevation differences stemming from bats' relative locations within a cave are often unpredictable or unmeasured, and (2) the site itself remains at the same elevation throughout the year. The current study groups Midwestern hibernacula by latitude into upper and lower Midwest groups rather than using elevation as a time-varying environmental node. The final environmental node, resources (Re), symbolizes energy sources available to bats living in the hibernaculum, which could include water and food both inside and outside the hibernaculum within reasonable flying distance. Standing water increases both humidity and available food, which includes various arthropod prey species such as flies, moths, stoneflies, and small beetles \cite{mccoshum_literature_2023}. Hence, Re can be considered predominantly driven by available water. This node affects all bats, as even healthy bats periodically arouse from torpor \cite{klug-baerwald_meet_2022}, but plays a substantially larger role for infected bats due to longer periods of arousal and a physiological stress response that more rapidly depletes fat reserves \cite{30days_immune, warnecke2012}. The second part of Table \ref{tab:nodes} summarizes these environmental nodes. The environmental reservoir used to represent fungal contamination is handled separately and is not an additional Boolean environmental node.

\begin{table}[H]
\centering
\renewcommand{\arraystretch}{1.2}
\begin{tabular}{>{\bfseries}l l p{8cm}}
\toprule
Category & Node & Description \\
\midrule

\multirow[t]{4}{*}{Inhabitants}
& Hibernating (Hi)
& Hibernating bats \\

& Immune (Im)
& Bats experiencing a period of heightened immunity following recovery from infection \\

& Infected (In)
& Non-hibernating bats infected with white-nose syndrome \\

& Other awake (Ot)
& Non-hibernating bats that are also not infected or immune, i.e. a susceptible population \\

& Deceased (De)
& Deceased bats \\

\midrule

\multirow[t]{5}{*}{Environment}

& Temperature (Te)
& Ambient temperature within the hibernaculum \\

& Humidity (Hu)
& Relative humidity within the hibernaculum \\

& Resources (Re)
& Standing water and food available to bats \\

\bottomrule
\end{tabular}
\caption{Model nodes categorized by ecological role and behavior status.}
\label{tab:nodes}
\end{table}

Within our node categories, an inhabitant state that is ON at a given time step indicates that the particular bat belongs to that subcategory. We represent each bat by an index $i$, with a state indicator for that bat in each inhabitant subcategory. Denote, for instance, $\text{Hi}_i(t)$ as the state indicator for bat $i$ in the hibernating subcategory at time $t$. If bat $i$ is hibernating, then $\text{Hi}_i(t)=1$, or ON, while $\mathrm{Im}_i(t)=\text{In}_i(t)=\text{Ot}_i(t)=\text{De}_i(t)=0$, or OFF. It is impossible for two subcategories corresponding to the same individual $i$ to be ON simultaneously. In fact, exactly one must be ON while the others are OFF. On the other hand, when an environmental node is ON, a particular environmental value has exceeded a threshold relevant to behavior such as hibernation. For instance, resources (Re) turns ON only when the amount of resources in the cave is sufficient to sustain bats during the day. In this way, a Boolean network model need not explicitly track the threshold values, only whether the relevant conditions have been met. Moreover, the fact that ON and OFF have different meanings for the different categories of nodes is reflected in how many nodes are in each category. There could be hundreds of hibernating bat state indicators (Hi), but there is exactly one temperature node (Te). All environmental nodes are unique and distinct from the inhabitants.

We note that each inhabitant acts as a parent node or input to the other inhabitants, meaning that the inhabitant network is \textit{fully connected}. For simplicity, we assume that every bat can interact with every other bat. Individuals in the same subcategory follow the same update rules, although their trajectories may differ because state transitions are stochastic. The update rules that govern the network will be presented in detail in Section \ref{subsec:rules}. For example, a bat awakening from hibernation is represented by its Hi indicator going from ON to OFF and its Ot indicator going from OFF to ON. Such transitions preserve the condition that each individual occupies exactly one inhabitant state; population turnover is accounted for separately in Section \ref{sec:model}.

\subsection{Model Parameters}\label{subsec:parameters}

Due to limited empirical data of WNS dynamics at the level of individual hibernacula, e.g. infection or mortality rates, we model key transition probabilities as random variables. This approach captures individual heterogeneity and environmental uncertainty, consistent with standard practices in epidemiology and ecological modeling \cite{anderson_infectious_1991,keeling2008}. The complete set of probabilities and threshold parameters is described in Table \ref{tab:python_params}. The core transition probabilities govern stochastic changes in bat states and population size. These probabilities are calculated from model parameters, some of which are sampled from probability distributions.

\renewcommand{\arraystretch}{1.1}
\begin{xltabular}{\textwidth}{l X}
\toprule
\textbf{Parameter} & \textbf{Description} \\
\midrule

\textbf{Core Transition Probabilities} & \textit{Calculated daily for individual bats or states} \\
$P_{\mathrm{inf},i}(t)$ & Probability that an ``Hi'' bat gets infected with WNS on the given day $t$ \\
$P_{\mathrm{dead},i}(t)$ & Probability that encompasses deaths from WNS infections and environmental pressure. $P_{\mathrm{dead},i}$ is determined by various strains on the bats, which can increase as injury from infection occurs and as resources deplete.\\
$P_{\text{recover}}$ & Probability of an ``In'' bat recovering and entering the temporary immune state (``Im'') on any given day \\
$P_{\mathrm{bouts}}^{\text{Hi}},\; P_{\mathrm{bouts}}^{\text{Ot}}$ & Probabilities of exiting hibernation (``Hi'' to ``Ot'') and returning to hibernation (``Ot'' to ``Hi''), respectively, during periodic bouts of arousal \\
$P_{\mathrm{seasonal}}(t)$ & Probability of a bat transitioning between ``Hi'' and ``Ot'' on day $t$ due to seasonal transition of ``Te'' \\
$P_{\mathrm{net}}(t)$ & Probability of adding a new bat through birth or immigration on day $t$, including during winter. \\

\midrule
\textbf{Environmental Dynamics} & \textit{Governing seasonal switches and spore decay} \\
$WS$ & Winter start: days after Jan. 1st that temperature node $Te$ turns off \\
$WL$ & Winter length: number of days $Te$ remains off \\
$\delta$ & Daily decay rate of $P. \text{destructans}$ spores on hibernaculum surfaces \\
$\lambda_{\text{win}}, \lambda_{\text{sum}}$ & Expected daily population turnover rates during winter and non-winter seasons \\

\midrule
\textbf{Infection \& Mortality Dynamics} & \textit{Governing transmission, disease progression, and strain} \\
$c_i$ & Cluster size of the $i$-th bat, scaling bat-to-bat contact rate $\lambda_i$ \\
$\alpha_{\mathrm{inf}}, \beta_{\mathrm{inf}}$ & Shape parameters of the Beta distribution governing per-contact transmission $p_i$ \\
$T_{\mathrm{inf}}$ & Average duration (days) an infected bat spends symptomatic before recovery or death\\
$k_{\mathrm{dead}}, \theta_{\mathrm{dead}}$ & Shape and scale parameters for the Gamma distribution of infection physiological strain $\mu_i$ \\

\midrule
\textbf{Arousal \& Seasonality Dynamics} & \textit{Governing torpor cycles and population transition timing} \\
$T_{\text{TBD}}, T_{\text{AD}}$ & Torpid bout duration and arousal duration (days), respectively \\
$T_{\mathrm{seasonal}}$ & Duration (days) of population transition in and out of hibernation when $Te$ switches \\

\midrule
\textbf{Immunity Dynamics} & \textit{Governing resistance: temporary, inherited, and acquired} \\
$r_i$ & Baseline individual resistance factor ($[0,1]$), affecting susceptibility to infection \\
$k_{\mathrm{im}}, \theta_{\mathrm{im}}$ & Shape and scale parameters of the Gamma distribution for temporary immunity duration $T_{\mathrm{im},i}$ \\
$T_{\mathrm{im},i}$ & Realized duration (days) individual $i$ spends in the temporary immune state ($\mathrm{Im}$) \\
$r_{\text{gain}}$ & Resistance gain multiplier permanently boosting $r_i$ upon exiting $\mathrm{Im}$ \\
$r_{\text{max}}$ & Standard deviation parameter for normally distributed inherited resistance in offspring  \\

\bottomrule
\caption{Structured parameters categorized by model mechanisms. Parameters specific to tricolored bats are: $\lambda_{\mathrm{win}}, \lambda_{\mathrm{sum}}, c_i$. Parameters specific to geographic location are: $\lambda_{\mathrm{win}}, \lambda_{\mathrm{sum}}, WS, WL$.}
\label{tab:python_params}
\end{xltabular}

\noindent More information about probability distributions and choices for the parameters that were not obtained from real data is given in Section \ref{sec:probabilitymodels}. These parameters are each addressed individually. We note that values for $c_i, T_{TBD},$ and $T_{AD}$ are approximated from ranges determined in literature found in Section \ref{sec:probabilitymodels}. We algorithmically approximate parameters that are not influenced by the presence of WNS or \textit{P. destructans} ($WS, WL, T_{\mathrm{seasonal}}, \lambda_{\mathrm{sum}}, \lambda_{\mathrm{win}}$) using a technique called Particle Swarm Optimization in Section \ref{sec:calibration}. The remaining parameters are only applied when WNS and/or \textit{P. destructans} are present. We cannot algorithmically approximate these parameters with the same accuracy due to a small amount of data, which is statistically over-dispersed due to the rapid dying-off of bat populations whenever WNS is contracted. Hence, we analyze the effect of these parameters in Section \ref{sec:analysis}.

\subsection{Boolean Update Rules}\label{subsec:rules}

Now that we have determined all relevant parameters, we define a discrete-time stochastic framework for WNS dynamics in bat hibernacula by defining rules that determine the time evolution of all inhabitants. All processes are formulated at the daily time scale using biologically motivated distributions. Each bat is updated through a set of coupled probabilistic transitions governing infection, mortality, recovery, behavior, population turnover, and environmental capacity constraints. To determine rules that update each bat in the population at successive time steps, we systematically determined biological links between nodes using ecological expert consultation and published literature \cite{simon_cave_2019, flory_environmental_2012, mccoshum_literature_2023, fortuna_roosting_2009}. \\

As indicated before, at each time step a bat occupies exactly one of the inhabitant subcategories described in Table \ref{tab:nodes}: Hi, In, Im, Ot, or De. The state of every bat, either 1 or 0, is updated from time $t$ to time $t+1$ according to Boolean rules that follow the biological behavior of each subcategory of inhabitants. We recall that the ON state means that the inhabitant under consideration is an existing member of that subcategory of the bat population. We start by describing the Boolean rules for the inhabitants, followed by the Boolean rules corresponding to the environment category. Recall we use the generic index $i$ to indicate the $i$-th individual in the category of inhabitants, also represented as $x_i$.\footnote{To simplify our notation, we use mathematical symbols such as $\wedge $ for the word ``AND" and $\vee $ for the word ``OR" in the Boolean formulas to avoid too many words. }


\subsubsection{Inhabitants}

\paragraph{Updating Hi.} The collection of Boolean rules that evolve the behavior of hibernating bats (Hi) is as follows:
    \begin{subequations}\label{eq:hi} 
    \begin{numcases}{\text{Hi}_i(t+1) =}
            1 \hbox{ if } \;\; \text{Ot}_i(t) = 1 \wedge  \text{Te}(t) = 0 \wedge \text{check}_i(t) = 0 \;\; \text{with probability }P_{\mathrm{seasonal}}(t),  \label{eq:hi:a} \\ 
            \;\; \hbox{ or} \;\; \text{Ot}_i(t) = 1  \wedge \text{Te}(t) = 0 \wedge \text{check}_i(t) = 1 \;\; \text{with probability }P_{\mathrm{bouts}}^{Ot},  \label{eq:hi:e}\\
            \;\; \hbox{ or} \;\; \mathrm{Hi}_i(t) = 1 \wedge \mathrm{Te}(t) = 1 \text{ with probability } 1-P_{seasonal}(t),   \label{eq:hi:c}\\
            \;\; \hbox{ or} \;\; \mathrm{Hi}_i(t) = 1 \wedge \mathrm{Te}(t) = 0 \text{ with probability } 1-P_{\mathrm{bouts}}^{Hi}, \label{eq:hi:b}\\ 
            \;\; \hbox{ or} \;\; \text{Im}_i(t) = 1 \wedge  T_{\mathrm{im},i} \leq 0 \label{eq:hi:d}, \\  
            0 \;\; \text{otherwise} \label{eq:hi:f} %
    \end{numcases}
    \end{subequations}

The first rule \eqref{eq:hi:a} represents a non-hibernating, susceptible bat, $\text{Ot}_i(t)$, that will go into hibernation in the next time step, $t+1$, due to decreasing temperature, Te$(t)= 0$. However, not all bats go into hibernation at the same time despite the lower temperature, so, under the right temperature conditions, a bat enters hibernation with probability $P_{\mathrm{seasonal}}(t)$. As mentioned earlier, the probability check can continue to fail, meaning the bat in the Ot state will continue to not enter hibernation in the next time step (this specific scenario is accounted for in the Updating Ot section). Similarly, the second rule \eqref{eq:hi:e} represents a non-hibernating, susceptible bat that goes into hibernation, but \textit{only after already awaking from hibernation temporarily}. To determine which case applies so that \eqref{eq:hi:a} and \eqref{eq:hi:e} are mutually exclusive, each bat $i$ tracks a Boolean variable $\text{check}_i(t)$ to determine whether they have already entered hibernation once in the winter season. If so, that bat is only eligible for \eqref{eq:hi:e}.

The third \eqref{eq:hi:c} and fourth \eqref{eq:hi:b} rules each represent when a bat is in the hibernation state, $\text{Hi}_i(t)=1$, and \textit{fails} to transition into another state. In \eqref{eq:hi:c}, the $i$th bat is already hibernating and $Te(t) = 1$, so it may fail the probability check that allows it to \textit{leave} hibernation, which is equivalent to passing the opposite probability check ($1- P_{\mathrm{seasonal}}$). Similar logic applies in the fourth rule, \eqref{eq:hi:b}, when the bat may experience a bout of arousal during wintertime and leave hibernation, corresponding with probability $P_{\mathrm{bouts}}^\text{Hi}$. So, in order to remain in the "Hi" state, the bat must pass the opposite probability check, ($1- P_{\mathrm{bouts}}^\text{Hi}$), which means that the bat will not become aroused and continue hibernating into the next time step. Note that periodic bouts of arousal are only features of bats in wintertime, 
    
The fifth rule \eqref{eq:hi:d} captures how an immune bat enters into hibernation. Since $T_{\mathrm{im},i}$ is decremented with each time step, a check is run to see if an immune bat has expended its temporary immune state. When the immune state is over, meaning when $T_{\mathrm{im},i} \leq 0$, then the bat automatically enters into hibernation.  

Any other scenario results in $\mathrm{Hi}_i(t+1)=0$, as specified in \eqref{eq:hi:f}. Similar reasoning is applied to each update rule function, meaning that any combinations of parental nodes not listed for a specific node update will always result in a 0 value in the next time step.

\paragraph{Updating In.} The collection of Boolean rules that evolve the behavior of bats infected with WNS (In) is as follows:
    \begin{subequations}\label{eq:in}
    \begin{numcases}{\text{In}_i(t+1) =} 
            1 \hbox{ if } \;\; \text{Hi}_i(t) =1 \wedge \text{Te}(t) =0, \;\; \text{with probability }P_{\mathrm{inf},i}(t) \label{eq:in:a} \\ 
            \;\; \hbox{ or} \;\; \text{In}_i(t) = 1 \wedge (\text{Te}(t) = 1 \vee \text{Re(t)} = 1 )  \label{eq:in:b} \\
            \qquad \;\; \text{with probability } 1 - P_{\mathrm{dead},i}(t) - P_{\mathrm{recover}} \notag
            \\ 
            0 \;\; \text{otherwise} \label{eq:in:c} %
    \end{numcases}
    \end{subequations}
 
The first rule \eqref{eq:in:a} models how a bat gets infected. Formally, a bat must be in hibernation, $\text{Hi}_i(t)=1$, and the temperature condition $\text{Te}(t)=0$ must be satisfied. Under these conditions, infection is assumed to occur with probability $P_{\mathrm{inf},i}(t)$.

The second rule \eqref{eq:in:b} models how an infected bat may stay infected at the next time step, first ruling out deterministic starvation while infected, i.e. requiring $\text{Te}(t)=1$ or $\text{Re}(t)=1$. Next, recovery and death are mutually exclusive with respective probabilities $P_{\mathrm{recover}}$ and $P_{\mathrm{dead},i}(t)$. Hence the probability of an infected bat failing both checks, each which transitions the bat into a different state, is given by $1 - P_{\mathrm{dead},i}(t) - P_{\mathrm{recover}}$, meaning that the bat will remain infected for the next time step. 
In all other cases, $\mathrm{In}_i(t+1)=0$, as specified in \eqref{eq:in:c}.

\paragraph{Updating Im.} The collection of Boolean rules that evolve the behavior of bats in a period of immunity (Im) after infection is as follows:
    \begin{subequations}\label{eq:im}
    \begin{numcases}{\mathrm{Im}_i(t+1) =} 
            1 \hbox{ if } \;\; \text{In}_i(t) = 1, \text{ with probability }P_{\text{recover}} \label{eq:im:a} \\ 
            \;\; \hbox{ or} \;\;\mathrm{Im}_i(t) =1 \wedge \;\text{$T_{\mathrm{im},i} > 0$} \label{eq:im:b} \\ 
            0 \;\; \text{otherwise} \label{eq:im:c} %
    \end{numcases}
    \end{subequations}
    
The first rule \eqref{eq:im:a} indicates how an infected bat, $\text{In}_i(t) =1$, can first enter the immune state, which is uniquely determined by probability $P_{\text{recover}}$. 
    
The second rule \eqref{eq:im:b} shows that a bat will remain in the immune state for a period of time governed by $T_{\mathrm{im},i}$. When a bat first becomes immune, $T_{\mathrm{im},i}$ is decreased every time step to track how long bats will have temporary immunity, and if there is still time remaining within a bat's immunity duration, i.e. $T_{\mathrm{im},i} > 0$, then the bat will be in the "Im" state in the next time step until $T_{\mathrm{im},i}$ becomes negative and the bat can re-enter hibernation, as seen in rule \eqref{eq:hi:d}. 
In all other cases, $\mathrm{Im}_i(t+1)=0$, as specified in \eqref{eq:im:c}.

\paragraph{Updating Ot.} The collection of Boolean rules that evolve the behavior of otherwise susceptible (Ot: not hibernating, immune, or infected) bats is as follows:
    \begin{subequations}\label{eq:ot}
    \begin{numcases}{\text{Ot}_i(t+1) =}
            1 \hbox{ if } \;\; \text{Hi}_i(t) = 1 \wedge \text{Te}(t) = 0, \;\; \text{with probability } P_{\mathrm{bouts}}^{\mathrm{Hi}} \label{eq:ot:a} \\
            \;\; \hbox{ or} \;\; \text{Hi}_i(t) = 1 \wedge \text{Te}(t) = 1, \;\; \text{with probability } P_{\mathrm{seasonal}}(t) \label{eq:ot:b} \\
            \;\; \hbox{ or} \;\; \big(\text{Ot}_k(t) = 1 \vee \mathrm{Im}_k(t) = 1\big) \text{ produces new individual } i, \;\; \text{with probability } P_{\mathrm{net}}(t) \label{eq:ot:c} \\
            \;\; \hbox{ or} \;\; \text{Ot}_i(t) = 1 \wedge \text{Te}(t) = 1 \label{eq:ot:d} \\
            \;\; \hbox{ or} \;\; \text{Ot}_i(t) = 1 \wedge \text{Te}(t) = 0 \wedge \text{Re}(t) = 1 \wedge \text{check}_i(t) = 1, \;\; \text{with probability } 1 - P_{\mathrm{bouts}}^{\mathrm{Ot}} \label{eq:ot:e} \\
            \;\; \hbox{ or} \;\; \text{Ot}_i(t) = 1 \wedge \text{Te}(t) = 0 \wedge \text{Re}(t) = 1 \wedge \text{check}_i(t) = 0, \;\; \text{with probability } 1 - P_{\mathrm{seasonal}}(t) \label{eq:ot:f} \\
            0 \;\; \text{otherwise} \label{eq:ot:g}
    \end{numcases}
    \end{subequations}

The first rule \eqref{eq:ot:a} indicates that during winter ($\text{Te}(t) = 0$), a hibernating bat $\text{Hi}_i(t) = 1$, experiences temporary and periodic bouts of arousal, with probability of this occurring on a given day equal to $P_{\mathrm{bouts}}^\text{Hi}$. Therefore, on the following time step, $\text{Hi}_i(t+1) = 0$ and $\text{Ot}_i(t+1) = 1$. The second rule \eqref{eq:ot:b} describes seasonal emergence when temperatures warm enough (Te$(t) = 1$) and the given bat enters the aroused state $\text{Ot}_i(t+1) = 1$ with probability $P_{\mathrm{seasonal}}(t)$. 

The third rule \eqref{eq:ot:c} focuses on population change, specifically resulting from births or immigration. Here, $k$ denotes an existing bat, while $i$ denotes a new individual entering the Ot state at time $t+1$.
The healthy (susceptible and immune) populations can produce offspring during the non-hibernation periods. Therefore, we can take a parent bat, either $\text{Ot}_k(t) = 1$ or $\text{Im}_k(t) = 1$, either of which produces an offspring with probability $P_{\mathrm{net}}(t)$. If successful, a new individual, $Ot_i(t+1)$, will be added into the Ot inhabitants list. In the code, this looks like appending a new element to represent the added bat. Notice that $\lambda_{\mathrm{net}}$ only adds bats to the population, as we assume an underlying trend of population growth irrespective of deaths due to WNS. All population declines thus occur via mortality in \eqref{eq:de}.

Rules \eqref{eq:ot:d}--\eqref{eq:ot:f} govern persistence in the active $\text{Ot}$ state. During warm weather ($\text{Te}(t) = 1$), active bats remain in the $\text{Ot}$ state, as the population is assumed to avoid infection while outside of hibernacula \eqref{eq:ot:d}. During winter ($\text{Te}(t) = 0$) and with resources available ($\text{Re}(t) = 1$), an aroused bat ($\text{check}_i(t) = 1$) remains active if it does not return to torpor, with probability $1 - P_{\mathrm{bouts}}^{\mathrm{Ot}}$ \eqref{eq:ot:e}, while a seasonally active bat ($\text{check}_i(t) = 0$) remains active if it does not yet enter hibernation, with probability $1 - P_{\mathrm{seasonal}}(t)$ \eqref{eq:ot:f}. In all other cases, $\mathrm{Ot}_i(t+1) = 0$ \eqref{eq:ot:g}.

\paragraph{Updating De.} The collection of Boolean rules that evolve deceased bats (De) is as follows:
    \begin{subequations}\label{eq:de} 
    \begin{numcases}{\text{De}_i(t+1) =}
            1 \hbox{ if } \;\; \text{Ot}_i(t) = 1 \wedge \text{Te}(t) = 0 \wedge \text{Re}(t) =0, \label{eq:de:a} \\ 
            \;\; \hbox{ or} \;\; \text{In}_i(t) =1 \wedge \text{Te}(t) = 0 \wedge \text{Re}(t) = 0, \label{eq:de:b} \\ 
            \;\; \hbox{ or} \;\; \text{In}_i(t) =1 \;\; \text{with probability }P_{\mathrm{dead},i}(t), \label{eq:de:c} \\ 
            \;\; \hbox{ or} \;\; \text{De}_i(t) = 1, \label{eq:de:absorbing} \\
            0 \;\; \text{otherwise} \label{eq:de:d} 
    \end{numcases}
    \end{subequations}

The first two rules in \eqref{eq:de} encapsulate deaths due to severe lack of resources ($\text{Re}(t) = 0$) during the winter ($\text{Te}(t) = 0$), induced by \textit{externally} forced factors such as regional drought. Together, these rules describe any non-hibernating bat (either Ot or In) dying from such resource depletion.
    
The third rule \eqref{eq:de:c} represents death as a result of physiological strain, potentially worsened by the lower level of resources inherent to a hibernaculum in wintertime, but ultimately distinct from \eqref{eq:de:a} and \eqref{eq:de:b} because it accounts solely for death due to \textit{internal} factors, especially as exacerbated by WNS. These deaths could be due to a bat's inability to fly effectively as a result of wing deterioration from an infection in the winter or to freezing from depletion of fat reserves or dehydration, even when moderate resources are present. However, not every infected bat will die, especially with the ability to replenish some of their fat reserves, so death occurs with probability $P_{\mathrm{dead},i}(t)$.
Rule \eqref{eq:de:absorbing} ensures that once a bat has died, it remains dead for the remainder of the simulation, provided deceased bats remain in the network. In all other cases, $\mathrm{De}_i(t+1)=0$, as specified in \eqref{eq:de:d}.

\subsubsection{\textbf{Environment}}

\paragraph{Updating Te.} The Boolean rule that models the temperature is as follows:
    \[
    \text{Te}(t+1) =
    \begin{cases}
    0 & \text{if } (t-WS)\bmod 365<WL,\\
    1 & \text{otherwise}.
    \end{cases}
    \]
Temperature, Te$(t)$, is only influenced by the current time step, $t$. Recall that $WL$ (winter length) is an integer that represents how many days a year Te$(t)$ is turned off, and $WS$ is an integer that represents the first day of the year that Te switches from on to off. Therefore, winter begins on day \(WS\) and lasts \(WL\) days, corresponding to \((t-WS)\bmod 365 < WL\). This pattern repeats every 365 days. At all other time steps, the temperature variable is set to 1, representing non-winter conditions.
   
\paragraph{Updating Re.} The Boolean rule that models the resources is as follows:
    \[
    \text{Re}(t+1) = \begin{cases}
        1 & \text{if } \text{Te}(t) = 1 \vee \text{Re}(t) = 1, \\
        0 & \text{otherwise.}
    \end{cases}
    \]
During warm weather, i.e. when Te$(t)= 1$, we assume that there are plenty of resources for bats, thus Re$(t+1) =1$, and in the winter time or colder climates, there would be a natural lack of resources, forcing the bat populations to go into hibernation.

\paragraph{Updating Hu.} The Boolean rule that models the humidity is as follows:
    \[
    \text{Hu}(t+1) = \begin{cases}
        1 & \text{if } \big(\text{Re}(t) = 1 \wedge \text{Te}(t) = 1\big) \vee \text{Hu}(t) = 1, \\
        0 & \text{otherwise.}
    \end{cases}
    \]
Similarly to Re, we consider Hu$(t+1)=1$ when weather is warm and moisture is present, i.e. Te$(t)=\text{Re}(t)=1$. Note that both Re and Hu are considered macro-environmental variables, so they are sustained as on throughout the simulations below, since we assume a baseline level of resources and humidity available in the winter. However, we retain this Boolean network architecture to support future investigation into regimes of severe drought and/or resource depletion. Moreover, a common concern of WNS is that lack of available resources in the winter exacerbate its severity. This \textit{individual} response of the $i$th bat to fat-reserve depletion and dehydration are captured in the physiological strain applied to the probability of mortality, as described in Section \ref{sec:death_param}.

\section{Parameterization and Computational Methods}\label{sec:methods}

To answer the questions posed by scientists, we constructed a Boolean network using the nodes of Section \ref{sec:nodes}. Next, we added a time-stepping mechanism that evolves the network over time according to the rules described in Section \ref{subsec:rules}, using the parameters introduced in Section \ref{subsec:parameters} and described in greater detail below. We recall that one time step represents a single day.

\subsection{Transition Probabilities and Parameter Specification}\label{sec:probabilitymodels}

In this subsection, we describe how probabilities are assigned to the major events in the Boolean network model as described in Table \ref{tab:python_params}. Since detailed ecological measurements are not available for every hibernaculum, we use biologically reasonable probability distributions commonly used in epidemiology, ecology, and network science \cite{anderson_infectious_1991,keeling2008}. Below, we define the function $\text{Count}$ as the total number of ON nodes in a given population. For example, $\text{Count}(\text{In}(t))$ is the total number of infected bats at time $t$. Moreover, we denote Alive to be the collection of all alive bats in the system so that 
\begin{equation*}
    \text{Count}(\text{Alive}(t)) = \text{Count}(\text{Hi}(t)) + \text{Count}(\text{In}(t)) + \text{Count}(\mathrm{Im}(t)) +\text{Count}(\text{Ot}(t)).
\end{equation*}

\subsubsection{Infection probability: $P_{\mathrm{inf},i}(t)$}\label{sec:inf_param}

We first define 
\begin{equation*}
    P_{\mathrm{inf},i}(t) = \Big(1 - \big(1 - P_{bat,i}(t)\big)\big(1 - P_{\mathrm{env}}(t)\big)\Big)(1-r_i),
\end{equation*}
to be the probability of a single bat getting infected with WNS at time $t$, where $P_{bat,i}$ represents infection due to contact from other bats and $P_{\mathrm{env}}$ represents infection due to contact with environmental reservoirs of \textit{P. destructans}, the fungus causing WNS. Also, $r_i \in [0,1]$ represents the resistance experienced by a given bat (discussed more in Section \ref{sec:imm_param}). \textbf{Note:} this probability applies to all Hi nodes when $\text{Te} =0$.

\paragraph{Infection from bat-to-bat contact: $P_{bat,i}(t)$}

Infection from other bats is modeled as a contact-driven stochastic process. The number of infectious contacts is 
$K_i(t) \sim \text{Poisson}(\lambda_i(t))$, where $\lambda_i(t)$ represents the expected daily infectious contact rate of one bat. This rate is determined by the size of group hibernating in close proximity to this bat. We call this group of bats associated with bat $i$ (including itself) a \textit{cluster} and denote its size $c_i$, so that 
\begin{equation*}
    \lambda_i(t) = (c_i -1) \frac{\text{Count}(\text{In(t)})} {\text{Count}(\text{Alive}(t))}.
\end{equation*}

When $\mathrm{Count}(\mathrm{Alive}(t))=0$, we set $\lambda_i(t)=0$, since no bats alive trivially implies no infections are occurring. For choices of clustering size $c_i$, we consider $c_i \in [1,10]$ sparse, $c_i \in [11,50]$ moderate, and $c_i\geq 51$ dense. For tricolored bats, $c_i \in [1,2]$ as they are not prone to clustering \cite{loeb_changes_2022}. For most bat species observed in North America, we consider clustering size $c_i \in [1,50]$  \cite{mccoshum_literature_2023}. As an example of another bat severely affected by WNS, big brown bats cluster in groups of 1-9 \cite{whitaker_hibernation_1992}, so one could take $c_i \in [1,9]$. Note that this coefficient $c_i$ varies both between and within species \cite{klys_bat_2024}.

Per-contact transmission probability is heterogeneous: $p_i \sim \text{Beta}(\alpha_{\mathrm{inf}},\beta_{\mathrm{inf}}),$ capturing variability in susceptibility and exposure. We resample this value at each time step, representing
daily variability in susceptibility and exposure. The resulting infection probability from bat-to-bat contact is $P_{bat,i}(t) = 1 - (1 - p_i)^{K_i(t)}$.
Ranges observed for varying beta distributions are consistent with contact-based wildlife transmission models and heterogeneous susceptibility assumptions in epidemic systems \cite{keeling2008}. Namely, we see that $\text{Beta}(1,10)$ (mean $\approx 0.09$) represents low transmission, $\text{Beta}(2,5)$ (mean $\approx 0.29$) represents moderate transmission, and $\text{Beta}(5,2)$ (mean $\approx 0.71$) represents high transmission.

\paragraph{Infection from environmental reservoirs: $P_{\mathrm{env}}(t)$}

Because \textit{P. destructans} has been shown to persist on hibernacula surfaces for long periods of time after complete extirpation of a bat population \cite{lorch_distribution_2013, hoyt_long-term_2015}, we model environmental fungal load as an environmental reservoir with constant decay rate per day $\delta$. The persistence -- and even propagation -- of \textit{P. destructans} depends on ambient hibernaculum temperature, as the fungus cannot grow in temperatures that exceed $20^\circ C$ \cite{hoyt_ecology_2021}. The exact lifespan of \textit{P. destructans} is unknown, but the fungus was found to persist over non-winter in 7 of 14 samples \cite{lorch_distribution_2013}, suggesting the ability to live for 1 to 2 years without hosts. Additionally, \textit{P. destructans} can live in laboratory conditions for at least 5 years \cite{hoyt_long-term_2015}, further validating a slow decay rate. While laboratory longevity is multi-year, field decay rates also incorporate seasonal microclimate turnover and ultraviolet exposure, supporting a slow decline during this time period. Therefore, we consider $\delta \in [0.005,0.03]$, corresponding to a half-life of approximately 23--138 days. This range represents varying rates of environmental fungal-load decline.

In \cite{lilley_landscape_2018}, Lilley et al. model the strain that different hibernaculum characteristics impose on bat populations affected by WNS using an SEI  partial differential equation model. In characterizing \textit{P. destructans} interaction with susceptible bats via environmental reservoirs, they use a sigmoidal relationship. We do the same and introduce an environmental contamination level $PD(t)$ to track fungal load as it decays at rate $\delta$:
\begin{align*}
    PD(t+1) = (1-\delta)PD(t) + \frac{\text{Count}(\text{In}(t))}{\text{Count}(\text{Alive}(t))}.
\end{align*}
When $\text{Count}(\text{Alive}(t))=0$, we also have $\text{Count}(\text{In}(t))=0$, so we simply set $PD(t+1) = (1-\delta)PD(t)$ as no bats are influencing the growth or decline of \textit{P. destructans}. We then define $P_{\mathrm{env}}(t) = PD(t) / \big( 1 + PD(t) \big)$, which is equivalent to a Hill function with exponent and weighting variable both equal to 1. This assumes a well-mixed hibernaculum in which environmental contamination arises from equal spore shedding by all infected bats.

\subsubsection{Mortality probability: $P_{\mathrm{dead},i}(t)$}\label{sec:death_param}

Mortality is modeled using a hazard formulation: $P_{\mathrm{dead},i}(t) = 1 - e^{-s_i(t)}$, which we define as the probability of a single bat dying from WNS infection at time $t$. \textbf{Note:} this probability applies to all In nodes.

Specifically, the individual hazard rate is controlled by $s_i$,
which combines the physiological and environmental strains
associated with WNS. We define $s_i(t) \coloneq \mu_i \rho(t)$,
broken up into two biological mechanisms. First, we define
the heterogeneous distribution
$\mu_i \sim \text{Gamma}(k_{dead},\theta_{dead})$,
where $\mu_i$ represents physiological strain from WNS.
This includes injury to wings and other soft tissues,
changes in thermoregulation, and more
\cite{30days_immune, warnecke2012}.
If a bat recovers from infection, it is known to do so
within approximately 30 days. In particular, bats which
do recover experience a rapid phase of tissue regrowth
during the second week post-infection, while inflammation
was still detectable up to day 40 \cite{30days_immune}.
Hence we can consider some expected time
$T_{\mathrm{inf}} \in [10,40]$ spent experiencing symptoms
of WNS infection before possible recovery.
Then we can define $\mu \coloneq 1/T_{\mathrm{inf}}
\in [0.025, 0.1]$, and we can let $k_{dead}=2$
and $\theta_{dead}=\mu/2$ to account for heterogeneity
in individual mortality hazards.
Gamma-distributed hazards are standard in survival
analysis for biological populations with heterogeneous
progression rates \cite{kleinbaum2012}.

While recovery and disease-induced mortality are distinct
physiological processes, both are parameterized using
$T_{\mathrm{inf}}$. This parameter represents the mean
waiting time for recovery in the absence of competing
mortality and also determines the scale of the
individual mortality hazard.

Next, we must take into account the additional environmental strain of resource depletion, making recovery from WNS even less feasible during winter. Hence we can define $\rho(t) = 2$ if $\text{Re}=0$ and $\rho(t)=1$ if $\text{Re} =1$, 
which signifies that low resource availability increases the mortality hazard associated with WNS.

\subsubsection{Recovery probability: $P_{recover}$}

Once a bat becomes infected, we utilize the fact that death and recovery are two mutually exclusive pathways based on the same time scale to simplify their respective formulations. \textbf{Note}: this probability applies to all In nodes.

Next, we write $P_{\text{recover}} = \min\{1/T_{\mathrm{inf}}, 1\}$, where $T_{\mathrm{inf}}$ represents the mean waiting time for recovery in the absence of competing mortality.
This memory-less recovery process, where the chance of recovery does not depend on how long the bat has already been infected, is simple yet reflects standard exponential waiting-time assumptions used in epidemiological compartment models \cite{keeling2008}.

\subsubsection{Transitioning between awake and torpid: $P_{\mathrm{bouts}}^{\mathrm{Hi}}, P_{\mathrm{bouts}}^{\mathrm{Ot}}$, $P_{\mathrm{seasonal}}(t)$}

Bats switch between awake and torpid under two circumstances: (1) when their internal energy stores run low enough that they need to scavenge for more food or when they experience a disturbance \cite{ben-hamo_waking_2013}, and (2) when seasons change and bats either enter or exit hibernation for the remaining winter or non-winter months, respectively. These transition probabilities apply to all Hi and Ot nodes.

\paragraph{Wintertime periodic bouts of arousal: $P_{\mathrm{bouts}}^{\mathrm{Hi}}, P_{\mathrm{bouts}}^{\mathrm{Ot}}$}\label{app:bouts}

During winter, bats periodically arouse from torpor before re-entering hibernation. 
Although arousal behavior is correlated with total evaporative water loss ($0.09 \leq R^2 \leq 0.20$ depending on local humidity) \cite{ben-hamo_waking_2013}, we do not explicitly model this relationship through the environmental nodes Re and Hu.
Hence we define additional parameters $T_{TBD}$ to be the mean torpor bout duration (days) before exiting hibernation, and $T_{AD}$ to be the mean arousal duration (days) before re-entering hibernation. Then the probabilities applied to Hi and Ot nodes are simply $P_{\mathrm{bouts}}^{\text{Hi}} = \min\{1/T_{TBD}, 1\}$ and $P_{\mathrm{bouts}}^{\text{Ot}} = \min\{1/T_{AD}, 1\}$, respectively. Notice that whenever $T_{TBD}$ or $T_{AD}$ are below 1, i.e. the corresponding bout duration lasts less than a day, we set the probability equal to 1 since our model is based on a daily time scale. In simulations, this means that a bat with $T_{AD}<1$ \textit{continues hibernating} unless otherwise disturbed.

Many factors of both the hibernaculum and individual bat -- like geographic location, propensity for disturbance, body condition, and more -- affect length of $T_{TBD}$ and $T_{AD}$. 
However, we use the reported mean durations as representative values for both torpor and arousal bouts.
For tricolored bats specifically, $T_{TBD} = 4.1 \pm 0.2$ days, while $T_{AD} = (88.5 \pm 46.7 \text{ minutes})/(1440 \text{ minutes/day})$ \cite{sirajuddin_winter_2025}. Note that because bats fall back asleep so quickly, they are not constrained by limited resources, unlike bats that exit hibernation due to infection. Moreover, because tricolored bats fall back asleep within the day, and our model is discretized to days, arousals shorter than one day are approximated as remaining within the same daily state. Hence, the transition probabilities
$P_{\mathrm{bouts}}^{\mathrm{Hi}}$ and
$P_{\mathrm{bouts}}^{\mathrm{Ot}}$ are capped at 1.
Arousal periods shorter than one day are
approximated as remaining within the same daily
state rather than producing separate transitions
between $\mathrm{Ot}$ and $\mathrm{Hi}$.

\paragraph{Seasonal transition: $P_{\mathrm{seasonal}}(t)$}

We model the probability that a bat transitions between active and hibernating states on a given day, driven by change in environmental temperature (\textit{Te}) between ON and OFF. Before entering and after exiting the hibernaculum, bats engage in swarming behavior that causes a gradual entrance and exit. In the fall, bats are observed to swarm and gradually enter the hibernaculum for about 1-1.5 months \cite{muthersbaugh_activity_2019}, while in spring, bats gradually exit the hibernaculum for about 30-50 days \cite{koch_how_2023}. Other than local weather and individual characteristics like sex, strongly correlative relationships with the exact emergence and exit time frames are unknown. 
Since local weather is outside the scope of this project, we use $T_{\mathrm{seasonal}}$ to represent the approximate duration of the transition between active and hibernating states following a seasonal change in temperature. Then we denote $P_{\mathrm{seasonal}}(t) = \min\{ 1, 1/(T_{\mathrm{seasonal}}-d)\}$, where $d$ is the number of elapsed days since the beginning of the current transition period. The value of $d$ resets to zero only at the onset of each seasonal shift and increments daily ($d = 0, 1, \dots, T_{\mathrm{seasonal}}-1$), creating an increasing daily transition hazard that ensures all bats complete the seasonal shift within $T_{\mathrm{seasonal}}$ days.

\subsubsection{Population turnover probability $P_{\mathrm{net}}(t)$}

Although hibernacula are relatively closed systems during winter, they are not perfectly isolated. Small-scale movement between nearby caves can occur, especially during disturbance events, and occasional mortality or migration unrelated to WNS may also affect observed counts. However, these events are rare during the hibernation period, making a rare-event stochastic model appropriate \cite{keeling2008,anderson_infectious_1991,russell2015}.

We model the number of population turnover events on day $t$ as a Poisson random variable,  $\Delta N(t)\sim\operatorname{Poisson}\!\left(\lambda_{net}(t)\right)$,
where the expected daily event rate depends on the seasonal state of the environmental temperature node,
\[
\lambda_{net}(t)=
\begin{cases}
\lambda_{\mathrm{win}}, & \text{if } Te(t)=0\\
\lambda_{\mathrm{sum}}, & \text{if } Te(t)=1.
\end{cases}
\]
Recall that $\lambda_{\mathrm{win}}$ and
$\lambda_{\mathrm{sum}}$ denote the expected
numbers of bats added to the population per
day during winter and non-winter, respectively.
With this implementation, we assume a baseline
healthy and non-migratory population that grows
over the course of years, with
$\lambda_{\mathrm{win}},\lambda_{\mathrm{sum}}>0$. The probability that at least one population turnover event occurs on day $t$ is therefore
\[
P_{\mathrm{net}}(t)
=
\mathbb{P}\!\left(\Delta N(t)\ge1\right)
=
1-\mathbb{P}\!\left(\Delta N(t)=0\right),
\]
which simplifies to
\[
P_{\mathrm{net}}(t)=
\begin{cases}
1-e^{-\lambda_{\mathrm{win}}}, & \text{if } Te(t)=0\\
1-e^{-\lambda_{\mathrm{sum}}}, & \text{if } Te(t)=1.
\end{cases}
\]

The bounds of parameters $\lambda_{\mathrm{win}}$ and $\lambda_{\mathrm{sum}}$ are estimated from the ecological literature and are summarized in Table \ref{tab:parameter_values}.

\subsubsection{Population immunity response}\label{sec:imm_param}

Immunity is the final major biological mechanism that drives population response to WNS infection. However, bat immunity response is perhaps the \textit{least} understood. Hence, we explore a combination of three different immune responses -- each with the ability to toggle in the Boolean network by setting relevant parameters to be trivial, hence enabling us to compare the different mechanisms. Each bat is paired with a resistance number $r_i \in [0,1]$, which is applied to $P_{\mathrm{inf},i}(t)$ (Section \ref{sec:inf_param}).

The first of these immune responses, and the most short-term, is achieved via the immune node Im, in which 
an infected bat ($\text{In}_i=1$) recovers ($\text{In}_i=0$ and $\text{Im}_i=1$). 
These bats experience temporarily higher immunity for $T_{im,i}$ days before being re-introduced into the susceptible hibernating population ($\text{Im}_i=0$ and $\text{Hi}_i=1$). This immune response is driven by the observation that the same immune suppression that makes bats vulnerable while hibernating is actually \textit{reversed} after re-entering hibernation post-infection \cite{meteyer_pathology_2012}. Moreover, recovery patterns of bats after surviving WNS infection provide evidence of a temporary immune period, since bats that survive WNS infection experience rapid clearing of fungal loads and re-growth of soft tissues in damaged wings \cite{30days_immune}. In the model, this period of immunity is achieved by setting $r_i=1$ for all Im bats. Upon return to Hi, bats may experience a more long-term, heightened resistance than before experiencing WNS, which is discussed more below. Additionally, we model $T_{im,i} \sim \text{Gamma}(k_{\mathrm{im}},\theta_{\mathrm{im}})$, rounded to the nearest integer, with shape parameter $k_{\mathrm{im}}$ and scale parameter $\theta_{\mathrm{im}}$, respectively \cite{keeling2008}. This allows for 0-day immunity, representing a bat that gains no substantial temporary resistance.

The next two immune responses are both long-term, since bat populations with high rates of WNS infection are known to stabilize -- or even grow -- several years after the initial outbreak \cite{langwig_resistance_2017, frank_evolution_2019}. For tricolored bats, we look at data from a South Carolina cave that experienced population stabilization after the third year post-infection and regrowth after the seventh \cite{south_carolina_tricolored}. This long-term transition from an epizootic to enzootic infectious regime leaves two possibilities: bats themselves experience some long-term resistance post-infection and offspring from bats once infected experience an increased resistance. To account for either possibility (or a combination of the two), we model both immune responses.

Long-term resistance of a bat that got infected and survived is achieved by increasing a bat's resistance number after exiting the immune class, Im. Since this distribution is entirely unknown, we introduce a simple multiplier $r_{\text{gain}} \in [0,1]$ that represents resistance gain. Suppose $r_i$ was the resistance number of a bat before entering Im, then upon exiting Im, we set $r_i = r_i + r_{\text{gain}}(1-r_i)$ (then set $r_i=\max\{0, \min\{1, r_i\}\}$ to avoid the case where $r_i \not\in [0,1]$) so that bats with high resistance do not experience much added benefit, while bats with low resistance experience more. 

The third and final of the immune responses is hereditary resistance, in which offspring of bats ``inherit'' their parents' immune response. In the model, bats enter the population by applying $P_{\mathrm{net}}(t)$ to a single bat $i$ on day $t$, so if we denote $r_i$ to be the parent's resistance to WNS, we can let the offspring's resistance $r_{\mathrm{offspring}}$ be equal to $r_i + N(0, r_{\mathrm{max}}^2)$ for some predetermined $r_{\mathrm{max}} < 1$. The resulting offspring resistance is then restricted to the interval $[0,1]$. Although bats entering the hibernaculum may be new immigrants, and hence have no resistance, our model does not discern incoming individuals by origin or prior exposure history, so every new bat added via $P_{\mathrm{net}}$ receives a resistance drawn from a Gaussian perturbation around some parent bat's own resistance \cite{barton_infinitesimal_2023}. Consequently, this implementation does not  accurately portray substantial influxes of migrant bats with no previous exposure to \textit{P. destructans}. Instead, it is most appropriate in species with high site fidelity, common to North American hibernating bats \cite{dixon_population_2011, norquay_long-distance_2013}.

\subsubsection{Summary}

In the previous sections, we described the parameters governing the stochastic dynamics and seasonal behavior of the model. Additionally, we have two time lengths $WL$ and $WS$, which describe the number of days spent in winter and the number of days after Jan. 1st that winter starts in the calendar year, respectively. Although $WS$ introduces an additional degree of freedom into the system, it allows the simulated winter period to be aligned with observations collected on different days of the calendar year, improving the accuracy of comparisons between simulations and observed data. Together, the stochastic dynamics of the system are governed by the following transition probabilities:
\begin{equation*}
    \{P_{\mathrm{inf},i}(t), P_{\mathrm{dead},i}(t), P_{\text{recover}}, P_{\mathrm{bouts}}^{\mathrm{Hi}},
P_{\mathrm{bouts}}^{\mathrm{Ot}}, P_{\mathrm{seasonal}}(t), P_{\mathrm{net}}(t)\},
\end{equation*}
along with adjustable parameters 
\begin{equation*}
    \{WL, WS, c_i, \alpha_{\mathrm{inf}}, \beta_{\mathrm{inf}}, T_{\mathrm{inf}}, T_{\mathrm{seasonal}}, T_{TBD}, T_{AD}, \delta, \lambda_{\mathrm{win}}, \lambda_{\mathrm{sum}}, 
    k_{\mathrm{im}}, \theta_{\mathrm{im}}, 
    r_{\mathrm{gain}}, r_{\max}\}.
\end{equation*}

This formulation provides a mechanistically interpretable stochastic framework that incorporates heterogeneity at the levels of contact structure, physiology, behavior, and environment, consistent with modern ecological disease modeling practice.

\subsection{Simulation Procedure}\label{sec:model}

To simulate the behavior of a population under a specific set of parameters, a different set of rules is applied in the winter and non-winter seasons. A pseudo-code layout of the simulation process is seen in Algorithm \ref{code:simulate} below.

\begin{algorithm}[H]
\caption{Hibernaculum Population Simulation}
\begin{algorithmic}[1]
\Procedure{Simulate}{}
    \State Initialize state (Hi, In, Im, Ot, De, Re, Te, Hu, PD)
    \For{$t = 1$ to $T$}
        \State Update season:
            $\textrm{Te} \gets
            \begin{cases}
            0, & \text{if } (t-WS)\bmod 365 < WL,\\
            1, & \text{otherwise}.
            \end{cases}$

        \State Record counts of (Hi, In, Im, Ot, De)
        \State $agg \gets$ aggregate(state) \textbf{// Count current populations} \\

        \State \textbf{// Environment update}
        \State Update (Re, Hu, PD) via logical rules on $(agg, Te)$ \\

        \State \textbf{// Population update}
        \For{each bat in Hi, In, Im, Ot}
            \State Apply probabilistic transitions:
            \State \quad $P_{\mathrm{inf},i}(t), P_{\mathrm{dead},i}(t), P_{\text{recover}}, P_{\mathrm{seasonal}}(t)$, $P_{\mathrm{bouts}}^{\mathrm{Hi}}, P_{\mathrm{bouts}}^{\mathrm{Ot}}$
            
            \State Add new bats with probability $P_{\mathrm{net}}(t)$, evaluated separately for each bat in Ot or Im
        \EndFor

        \State Update state
    \EndFor
    \State \Return history
\EndProcedure
\end{algorithmic}
\label{code:simulate}
\end{algorithm}

However, we have several unknown  parameters involved in the \textbf{Population update} section of Algorithm \ref{code:simulate} that prevent us from accurately predicting general population numbers. We are especially interested in finding the likelihood of a tricolored bat getting infected, dying, or recovering to inform the survivability of bats in a specific hibernaculum after introduction to WNS. This will inform the long-term survivability of a bat population. To determine the effects of these parameters of interest, we will first narrow down the remaining unknowns.

\subsection{Data and Parameter Estimation}\label{sec:calibration}

To calibrate our model and identify realistic values for the parameters in Table~\ref{tab:python_params}, we compare model predictions with observed population data from hibernacula across the Midwestern United States. Specifically, we focus on estimating parameters that do not directly govern a population's response to WNS.

\subsubsection{Data Selection}

We utilized data archived in the North American Bat Monitoring Program (NABat) \cite{NABat_data}, specifically looking at sites across the twelve Midwestern states with populations untouched by WNS. We filtered the data we received by (a) for every site, excluding years after the first instance of population decline at that site \textit{on or following} the first documented case of WNS in the corresponding state\footnote{2010 in Ohio \cite{WNS_ohio}; 2011 in Indiana \cite{WNS_indiana}; 2013 in Illinois \cite{WNS_illinois}; 2014 in Wisconsin \cite{WNS_wisconsin}}, which aims to exclude population declines associated with WNS. After removing data from those years, we included only sites that (b) had at least five years of usable data and (c) had more than 10 bats for most of those years to prevent high statistical dispersion of data, hence increasing the potential accuracy of parameter estimation. Of the data received, filtering with these methods narrowed down our data to thirty-five sites across four states:

\begin{figure}[H]
    \centering
    \includegraphics[width=\linewidth]{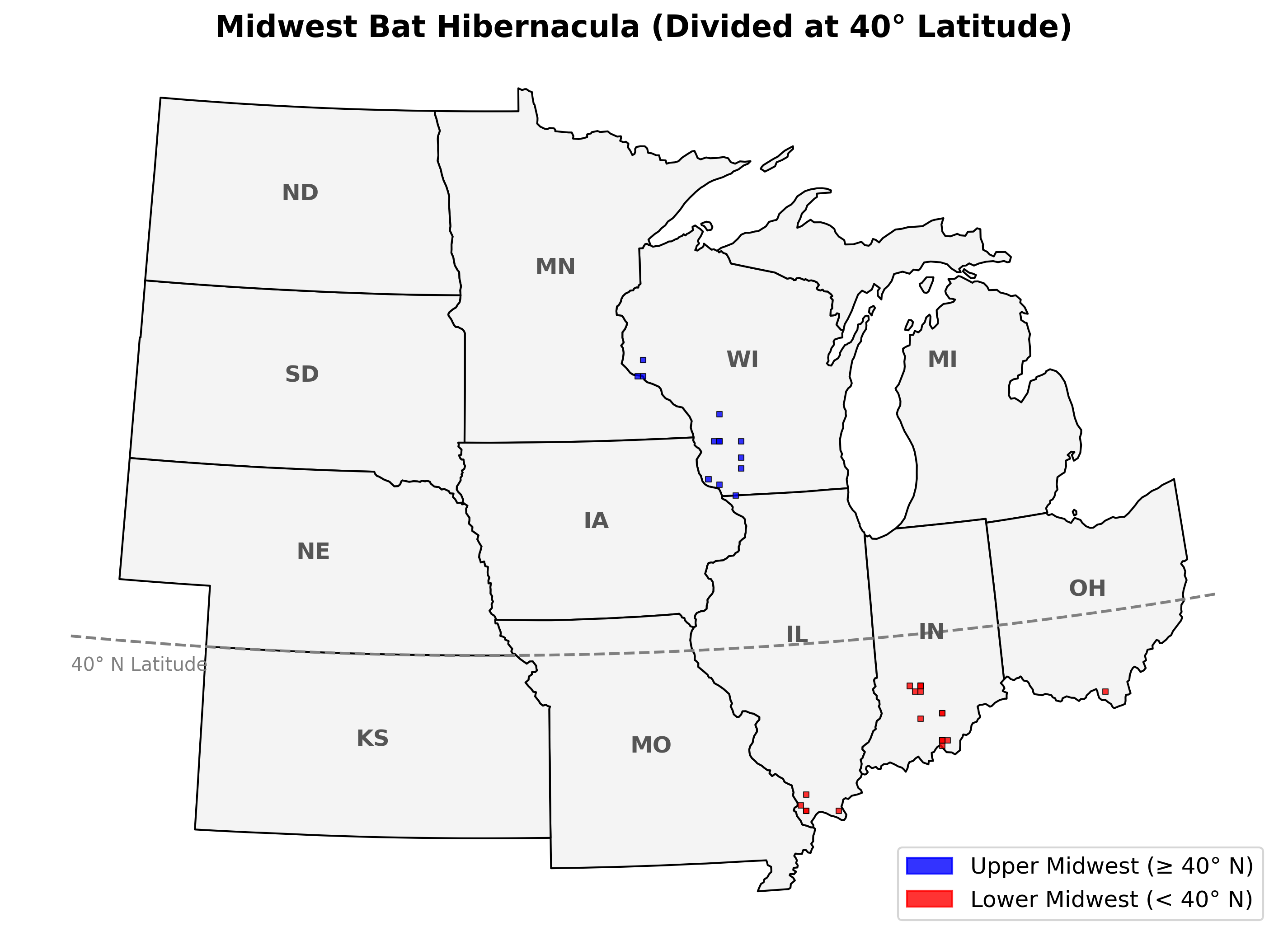}
    \caption{Grouping of sites across the Midwest}
    \label{fig:midwest_sites}
\end{figure}
Since the sites are cleanly split between the North and South Midwest, we divide them into two groups above and below the latitude $40^\circ$ North. For brevity, we call these groupings North Midwest and South Midwest, respectively.

\subsubsection{Parameter Estimation}

Since we are approximating parameters using only data from populations not reached by WNS, we can set initial number of infected bats in the model equal to 0, then approximate any parameters that still apply, e.g. parameters used to calculate $P_{\mathrm{net}}(t)$ but not $P_{\mathrm{inf},i}(t)$, based on recorded data. Specifically, we estimate the time scales $T_{\mathrm{seasonal}}$, $WS$, and $WL$, along with parameters associated with net change fluctuations, $\lambda_{\mathrm{win}}$ and $\lambda_{\mathrm{sum}}$. Of the remaining parameters, we recall that $c_i, T_{TBD},$ and $T_{AD}$ are estimated from the literature. The final eight remain unknown but need not be estimated at this stage because they have no influence over populations when there are no WNS cases and \textit{P. destructans} is not present. These eight are $\delta, \alpha_{\mathrm{inf}}, \beta_{\mathrm{inf}}, T_{\mathrm{inf}}, k_{\mathrm{im}}, \theta_{\mathrm{im}}, r_{\mathrm{gain}}, r_{\mathrm{max}}$. Once estimated, we set each of the values for $T_{\mathrm{seasonal}}$, $WS$, $WL$, $\lambda_{\mathrm{win}}$, and $\lambda_{\mathrm{sum}}$ for each of the simulations performed in Section \ref{sec:analysis}. Unless otherwise specified, we set each of the remaining eight parameters equal to the approximate mean of the possible values that parameter may take on, which we determine from literature (Table \ref{tab:sobol_parameter_values}).

Some parameter values were selected from published
studies whenever direct measurements existed
($T_{\mathrm{AD}}$ and $T_{\mathrm{TBD}}$).
For parameters not explicitly reported in the
literature, biologically plausible baseline
values and ranges were established using
published ecological information.
These values guided the selection of bounds for
Particle Swarm Optimization (PSO) calibration.
The resulting PSO-calibrated parameter values
were used in the final simulations.

To get approximate values, we use the method of PSO, as shown in Algorithm \ref{code:pso}. This algorithm is run over $2^8$ different sets of parameters, called \textit{particles}, which can be viewed as points of the 5-dimensional parameter space, where the dimensions are $T_{\mathrm{seasonal}}, WS, WL, \lambda_{\mathrm{win}},$ and $\lambda_{\mathrm{sum}}$ spaces. Additionally, we restrict this 5-dimensional parameter space to a bounded subset given by some bounding set $\mathcal{B}$ based on approximate ranges of parameters. For instance, we expect the onset of winter, represented by $WS$, to lie somewhere around the 300th day of the calendar year, as in Table \ref{tab:parameter_values}. Casting a wider net for potential variations, we use the range $WS \in [200,350]$ to determine a precise estimate for $WS$ using Algorithm \ref{code:pso}. So, the section of $\mathcal{B}$ corresponding with parameter $WS$ will consist of particles randomly selected in the interval $[200,350]$. Similar bounds for $\mathcal{B}$ are developed for each of these five parameters based on realistic baselines derived from ecological literature, as summarized in Table \ref{tab:parameter_values} below. In general, we say parameter $j$ is sampled in $[l_j, u_j]$. 

\begin{algorithm}[H]
\caption{Parallel Particle Swarm Optimization (PSO) for Parameter Estimation}
\begin{algorithmic}[1]
\Procedure{PSO}{$N, K_{\text{max}}$, $\mathcal{B}$, $P$}
    \State Initialize $N$ particles with random parameter vectors $\mathbf{x}_i \in \mathcal{B}$ and velocities $\mathbf{v}_i$
    \State Set personal bests $\mathbf{p}_i \gets \mathbf{x}_i$ and global best $\mathbf{g} \gets \arg\min_{\mathbf{p}_i} \text{loss}(\mathbf{p}_i)$

    \For{iteration $k = 1$ to $K_{\text{max}}$}
        \State $w \gets 0.9 - 0.5(k / K_{\text{max}})$ \Comment{Linear inertia decay}

        \State \textbf{Parallel Evaluation (Message Passing Interface, MPI):}
        \For{each particle $i = 1$ to $N$ \textbf{in parallel}}
            \State Simulate model using parameter set $\mathbf{x}_i$
            \State Compute loss $\ell_i$ (Negative binomial 2 negative log-likelihood against observed data)
            \If{$\ell_i < \text{loss}(\mathbf{p}_i)$}
                \State $\mathbf{p}_i \gets \mathbf{x}_i$ \Comment{Update personal best}
            \EndIf
        \EndFor

        \State Update global best $\mathbf{g}$ to the overall lowest loss across all $\mathbf{p}_i$

        \State \textbf{Swarm Movement, Bounding, \& Mutation:}
        \For{each particle $i = 1$ to $N$}
            \State Sample random vectors $\mathbf{r}_1, \mathbf{r}_2 \sim \mathcal{U}(0, 1)$
            \State $\mathbf{v}_i \gets w \mathbf{v}_i + c_1 \mathbf{r}_1 \odot (\mathbf{p}_i - \mathbf{x}_i) + c_2 \mathbf{r}_2 \odot (\mathbf{g} - \mathbf{x}_i)$
            \State $\mathbf{x}_i \gets \mathbf{x}_i + \mathbf{v}_i$

            \If{$\mathbf{x}_i$ violates boundaries $\mathcal{B}$}
                \State Clamp $\mathbf{x}_i$ to boundary limits and set $\mathbf{v}_i \gets -0.5 \mathbf{v}_i$ \Comment{Dampened reflection}
            \EndIf

            \If{$\text{rand}() < P$} \Comment{Mutation chance}
                \State Resample $\mathbf{x}_i$ and $\mathbf{v}_i$ uniformly at random within $\mathcal{B}$
            \EndIf
        \EndFor
    \EndFor

    \State \Return Best parameter vector $\mathbf{g}$
\EndProcedure
\end{algorithmic}
\label{code:pso}
\end{algorithm}

\begin{table}[H]
\centering
\small
\renewcommand{\arraystretch}{1.1}
\begin{tabular}{p{2.0cm} p{1.9cm} p{2.2cm} p{6.5cm}}
\hline
\textbf{Parameter} & \textbf{Baseline} & \textbf{Range} & \textbf{Basis for selection} \\
\hline
$WS$ & 300th day of the year & $[200,350]$ &
Calendar alignment between simulation and seasonal observations. The range allows variation in synchronization dates \cite{muthersbaugh_activity_2019,mccoshum_literature_2023}. \\
$WL$ & 150 days & $[120,240]$ days &
Simplified winter duration informed by reported hibernation and emergence patterns; the range allows seasonal variation \cite{mccoshum_literature_2023,muthersbaugh_activity_2019}. \\
$\lambda_{\mathrm{sum}}$ & 0.01 chance per day & $[0.0001,0.0004]$ chance per bat &
Assumed non-winter population-change rate, giving an approximately $1\%$ daily event probability. The range is exploratory, informed by seasonal activity \cite{mccoshum_literature_2023}. \\
$\lambda_{\mathrm{win}}$ & 0.001 chance per day & $[0,0.0001]$ chance per bat &
Assumed winter population-change rate, giving an approximately $0.1\%$ daily event probability. The lower rate reflects reported hibernaculum fidelity \cite{mccoshum_literature_2023}. \\
$T_{\mathrm{seasonal}}$ & 30 days & $[40,80]$ &
Assumed transition duration reflecting uncertainty in autumn and spring activity patterns \cite{muthersbaugh_activity_2019,mccoshum_literature_2023}. \\
\hline
\end{tabular}
\caption{Baseline parameter values and ranges used in the Boolean network simulations.}
\label{tab:parameter_values}
\end{table}

Notice that our ranges for $\lambda_{\mathrm{sum}}$
and $\lambda_{\mathrm{win}}$ are significantly
lower than our preliminary 
baseline values. The baselines were estimated
at the population level, whereas our model
evaluates population turnover separately for
each bat on a daily basis. The ranges in
Table~\ref{tab:parameter_values} reflect
this distinction and provide preliminary
bounds for parameter estimation.

Similar bounds are established for the other parameters estimated. At each of the $K$ iterations of PSO (for some $K \leq K_{\text{max}}=1000$, determined based on sufficient stopping conditions\footnote{The algorithm stops if the best loss has not improved upon the global loss by at least machine precision $(10^{-16})$ for 30 consecutive iterations}), the position of each particle is updated according to its velocity, which depends on its previous velocity, its personal best position, and the global best position. An initial velocity associated with parameter $j$ is sampled uniformly in $[-0.1(u_j-l_j), 0.1(u_j-l_j)]$. Cognitive and social coefficients are both set to the standard benchmark value of $c_1=c_2=1.49618$, while the inertia weight $w$ decreases linearly from $0.9$ to $0.4$ over the iterations, reducing the influence of the previous velocity as the optimization progresses. These choices are standard for balancing exploration across parameter bounds and convergence around high-likelihood regions.

Finally, to reduce the risk of the solution getting stuck in an arbitrary local minimum -- which itself may be very far from the global minimum -- we introduce a small probability ($P = 0.05$) called the \textit{mutation chance}  of any particle getting resampled anywhere within the bounds $B$. More detailed information about PSO can be found in \cite{kennedy_particle_1995}.

The objective function minimized by PSO in Algorithm
\ref{code:pso} is the negative log-likelihood (NLL),
a standard approach in ecological modeling
\cite{bolker_ecological_2008,hilborn_ecological_1997,king_avoidable_2015}.
We assume that observed bat counts follow a
negative binomial 2 (NB2) distribution around
model-predicted counts. This distribution
accommodates overdispersion commonly observed
in ecological population data, with mean $\mu$
and variance $\sigma^2=\mu+\mu^2/r$, where $r$
is the dispersion parameter
\cite{cameron_regression_1998}.
Minimizing the NLL allows PSO to account for
variability in observed counts without attempting
to reproduce every fluctuation in the data.
The dispersion parameter $r$ is also optimized
during parameter estimation, where it is treated as an additional component of each PSO particle and sampled in the bounds $[0.01,100]$.

This simulation is run for all thirty-five hibernacula
in our data set, creating thirty-five sets of calibrated
parameters. Although we already restrict populations
used in our data to more than ten bats for most observation years to prevent
overly small data sets that lead to high variance in
observed counts, we also use the following initialization protocol for the simulations in Algorithm~\ref{code:pso}:
for any given data set to which we calibrate our
parameters, the initial population for every simulation
used in Algorithm~\ref{code:pso} is taken to be the
number of bats in the very first observation, which
we consider to be healthy and awake (Ot) tricolored bats.

\begin{figure}[H]
    \centering
    \includegraphics[width=\linewidth]{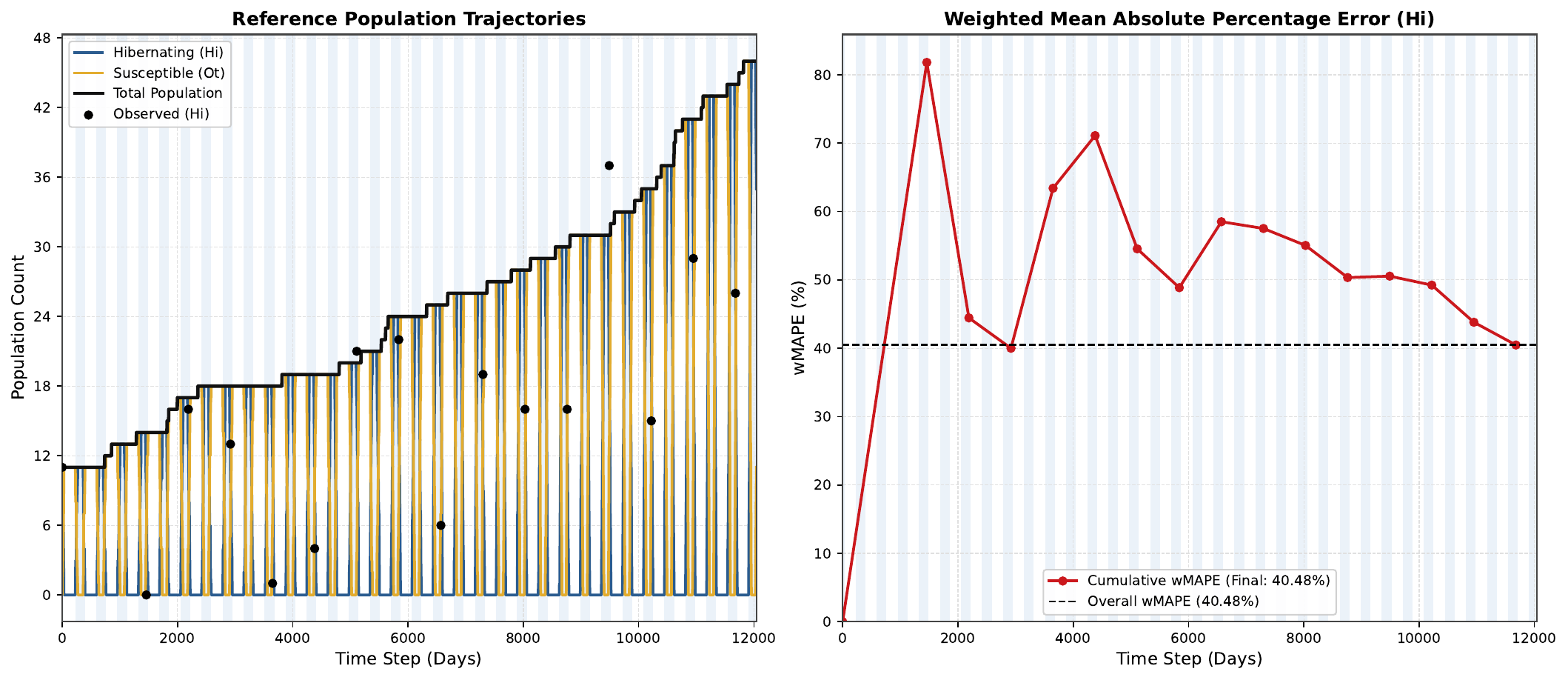}
\caption{Parameter optimization results for the Batwing Cave in Indiana, obtained using PSO}
    \label{fig:training}
\end{figure}

Figure~\ref{fig:training} illustrates the
results of PSO parameter calibration for
Batwing Cave in Indiana.

In Figure~\ref{fig:training}, the shaded blue regions
indicate winter periods. The blue and orange curves
represent the simulated hibernating (Hi) and susceptible
(Ot) populations, respectively, while the thicker black
curve represents the total simulated population. Black
dots indicate observed hibernating bat counts. The
simulation captures the overall increasing population
trend observed in the data, together with the seasonal
transitions between hibernating and active states.
Although the simulated counts generally exceed the
observations, the model reproduces the broader
population trajectory despite considerable variability
in the field measurements.

To quantify the agreement between observed and simulated
hibernating populations, the right panel shows the
cumulative weighted mean absolute percentage error
(wMAPE), defined at time $t$ as
\[
\mathrm{wMAPE}(t)
=
\frac{\displaystyle\sum_{j:\,t_j\leq t}
|x_{\mathrm{obs}}(t_j)-x_{\mathrm{sim}}(t_j)|}
{\displaystyle\sum_{j:\,t_j\leq t}x_{\mathrm{obs}}(t_j)}
\times 100\%.
\]
Here, $t_j$ denotes the time of the $j$-th observation,
while $x_{\mathrm{obs}}(t_j)$ and $x_{\mathrm{sim}}(t_j)$ denote the
observed and simulated hibernating bat counts,
respectively. Following an initial increase and
subsequent fluctuations, the cumulative wMAPE
generally declines toward the end of the simulation,
reaching $40.48\%$. This pattern indicates that
the cumulative relative discrepancy becomes smaller
over the later portion of the simulation. Differences
between simulated and observed counts may partly
reflect year-to-year variation in hibernation timing,
bat movements before surveys, and uncertainty in
field counts. Overall, these results illustrate the
model's ability to reproduce the broad population
trend while accommodating seasonal dynamics.

Finally, as a result of these computations, we estimate that after rounding to the nearest whole number, the set of parameters $\{T_{\mathrm{seasonal}}, WS, WL\}$ has an average of $\{63, 288, 172\}$ for North Midwest sites and $\{55, 280, 160\}$ for South Midwest sites. The set of parameters $\{\lambda_{\mathrm{sum}}, \lambda_{\mathrm{win}}\}$ has an average of $\{3.05 \times 10^{-4}, 6.0 \times 10^{-6}\}$ for North Midwest sites and $\{3.25 \times 10^{-4}, 5.0 \times 10^{-6}\}$ for South Midwest sites. The estimated parameter values are generally similar
between the two regions. The estimated $T_{\mathrm{seasonal}}$ is longer for
North Midwest sites (63 vs.\ 55 days), indicating a longer seasonal transition period in the model. Although we only consider bands of latitude, some discrepancies between our northern and southern sites may be due to the karst cave system of Indiana being the terrain where most of the southern Midwest caves were sampled (e.g. Batwing, Buckner, Ray's, Big Wyandotte, and more) \cite{mumford_mammals_1982}. These caves act as large subterranean cavities that
trap heat, cooling down in the beginning of October
and not warming again until late March
\cite{richter_modified_1993}. Such local cave
conditions may help explain why the estimated $WL$
values are relatively close (172 days in the north
and 160 days in the south), despite differences
in regional surface climates.

\section{Model Analysis and Results}\label{sec:analysis}

Next, we investigate the effects of the remaining
unknown parameters on the total population of
tricolored bats. Recall that for tricolored bats,
we consider $c_i \in [1,2]$ for each bat $i$,
which we implement as a uniformly random distribution.
At this point, the original set of unknown parameters
has been reduced to eight:
$\{\delta, \alpha_{\mathrm{inf}}, \beta_{\mathrm{inf}},
T_{\mathrm{inf}}, k_{\mathrm{im}}, \theta_{\mathrm{im}},
r_{\mathrm{gain}}, r_{\mathrm{max}}\}$.

\subsection{Parameter Sensitivity Analysis}

To quantify the model's sensitivity to variations in the unknown parameters, we performed a variance-based global sensitivity analysis using Sobol' indices \cite{Sobol_1990aa}. We evaluated our two sets of latitudinal parameters -- representing the North Midwest ($\ge 40^\circ\text{N}$) and South Midwest ($< 40^\circ\text{N}$) -- across four key epidemiological metrics: the empirical basic reproduction number ($R_0$), peak daily prevalence ($P_{\mathrm{max}}$), final population persistence ($S_{\mathrm{final}}$), and final cumulative mortality ($M_{\mathrm{final}}$).

Given a nonzero number of initially infected bats, we define $R_0$ as the number of secondary infections divided by the initial number of infected bats. We calculate prevalence via $P(t) \coloneq \text{Count}(\mathrm{In}(t))/ \text{Count}(\mathrm{Alive}(t))$; we calculate persistence normalized by the initial population as $S(t) \coloneq \text{Count}(\mathrm{Alive}(t))/\text{Count}(\mathrm{Alive}(0))$; and we calculate cumulative mortality, similarly normalized, as $M(t) \coloneq \text{Count}(\mathrm{De}(t))/\text{Count}(\mathrm{Alive}(0))$. Then $P_{\mathrm{max}} \coloneq \max_t \{P(t)\}$, where the max is taken over all simulated days $t$. Also, $S_{\mathrm{final}} \coloneq S(T)$ and $M_{\mathrm{final}} \coloneq M(T)$. For these simulations, we let $T=365 \times 20$, as we later find that populations settle into a steady state closer to the decadal scale.

Using the quasi-random sampling scheme designed for Sobol' analysis, called Saltelli sampling, we generated $N=2^{12}$ base samples. Again, we used ranges motivated by standard values each of these parameters may take in biological and epidemiological literature, which are defined in Table \ref{tab:sobol_parameter_values} below. Each Saltelli parameter combination was simulated ten
times, and the four output metrics were averaged
across these runs before computing the Sobol' indices. Also, remaining parameters ($WS, WL, T_{\mathrm{seasonal}}, \lambda_{\mathrm{win}}, \lambda_{\mathrm{sum}}, c_i, T_{\mathrm{TBD}}, T_{\mathrm{AD}}$) were assigned their calibrated or literature-based values, discussed in Section \ref{sec:calibration}.

\begin{table}[H]
\centering
\small
\renewcommand{\arraystretch}{1.1}
\begin{tabular}{p{2.0cm} p{1.9cm} p{2.2cm} p{6.5cm}}
\hline
\textbf{Parameter} & \textbf{Baseline} & \textbf{Range} & \textbf{Basis for selection} \\
\hline
$\alpha_{\mathrm{inf}}$ & $2$ & $[1, 5]$ &
First shape parameter of $\mathrm{Beta}(\alpha_{\mathrm{inf}}, \beta_{\mathrm{inf}})$ per-contact transmission; spans low-to-high contact infectivity regimes standard in heterogeneous wildlife disease models \cite{keeling2008}. \\
$\beta_{\mathrm{inf}}$ & $5$ & $[2, 10]$ &
Second shape parameter of $\mathrm{Beta}(\alpha_{\mathrm{inf}}, \beta_{\mathrm{inf}})$; paired with $\alpha_{\mathrm{inf}}$, yields mean per-contact transmission probabilities from $\approx 0.09$ ($\mathrm{Beta}(1,10)$) to $\approx 0.71$ ($\mathrm{Beta}(5,2)$) \cite{keeling2008}. \\
$\delta$ & $0.01\text{ day}^{-1}$ & $[0.005, 0.05]$ &
Daily environmental spore decay rate corresponding to fungal half-lives of $\approx 14$--$138$ days, capturing multi-month substrate persistence across summer dormancy up to faster microclimate degradation \cite{lorch_distribution_2013,hoyt_long-term_2015}. \\
$T_{\mathrm{inf}}$ & $30\text{ days}$ & $[10, 40]\text{ days}$ &
Mean symptomatic infection duration prior to recovery or mortality, grounded in empirical observations of wing-tissue regrowth within two weeks and residual inflammation up to 40 days \cite{30days_immune}. \\
$r_{\mathrm{max}}$ & $0.1$ & $[0, 0.8]$ &
Standard deviation of inherited resistance variation around parental values, 
spanning no random variation in inherited resistance ($0$) to substantial variation ($0.8$) within the unit resistance scale.
 \\
$r_{\mathrm{gain}}$ & $0.2$ & $[0, 1]$ &
Fractional post-recovery resistance gain multiplier; spans the full theoretical unit interval from no lasting protection ($0$) to complete acquired resistance ($1$). \\
$k_{\mathrm{im}}$ & $2$ & $[0.1, 10]$ &
Shape parameter of the Gamma temporary immunity distribution, spanning highly right-skewed immunity durations ($k_{\mathrm{im}} < 1$) to more concentrated durations ($k_{\mathrm{im}} = 10$).
 \\
$\theta_{\mathrm{im}}$ & $15\text{ days}$ & $[10^{-15}, 30]\text{ days}$ &
Scale parameter of the Gamma temporary immunity distribution; combined with $k_{\mathrm{im}}$, spans effectively immediate waning through multi-month seasonal protection consistent with post-hibernation immune waning \cite{30days_immune}. \\
\hline
\end{tabular}
\caption{Baseline values and sampling bounds for the eight unconstrained epidemiological parameters evaluated in the global Sobol' sensitivity analysis.}
\label{tab:sobol_parameter_values}
\end{table}

We measure sensitivity using first-order ($S_1$) and
total-order ($S_T$) Sobol' indices. First-order
indices measure the fraction of output variance
attributable to each parameter individually,
while total-order indices account for both its
individual contribution and its interactions
with other parameters.

\begin{figure}[H]
  \centering
  
  \begin{subfigure}[b]{0.49\textwidth}
    \centering
    \includegraphics[width=\textwidth]{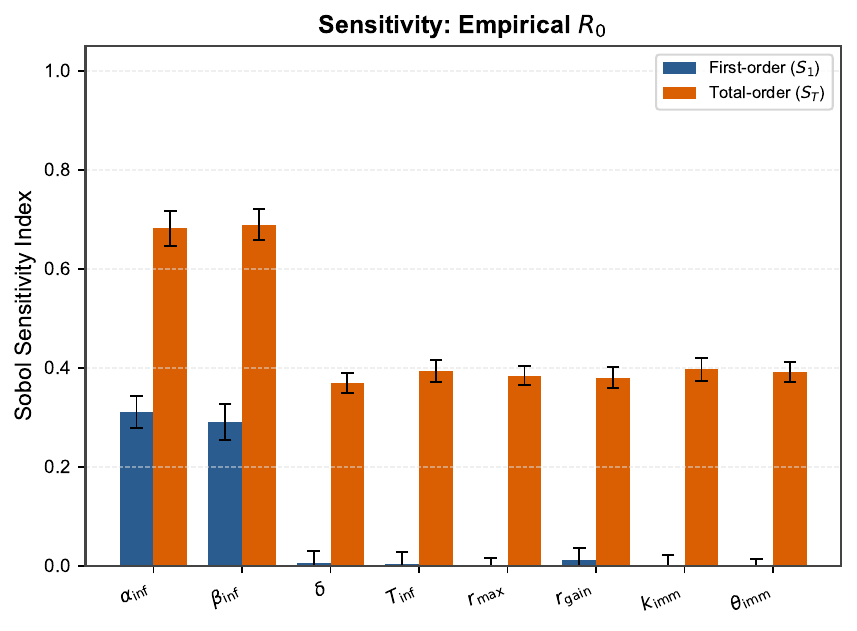}
  \end{subfigure}
  \hfill
  \begin{subfigure}[b]{0.49\textwidth}
    \centering
    \includegraphics[width=\textwidth]{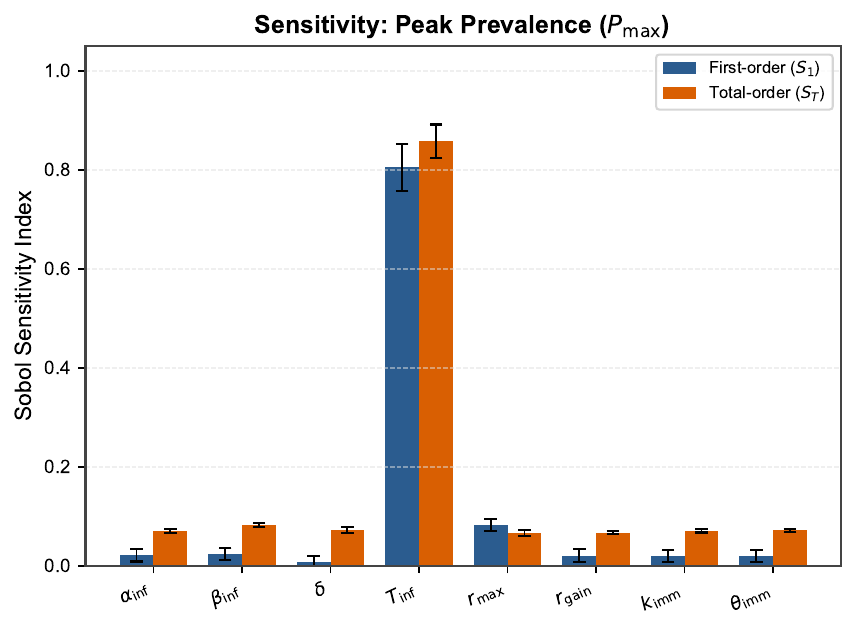}
  \end{subfigure}
  
  \medskip 
  
  \begin{subfigure}[b]{0.49\textwidth}
    \centering
    \includegraphics[width=\textwidth]{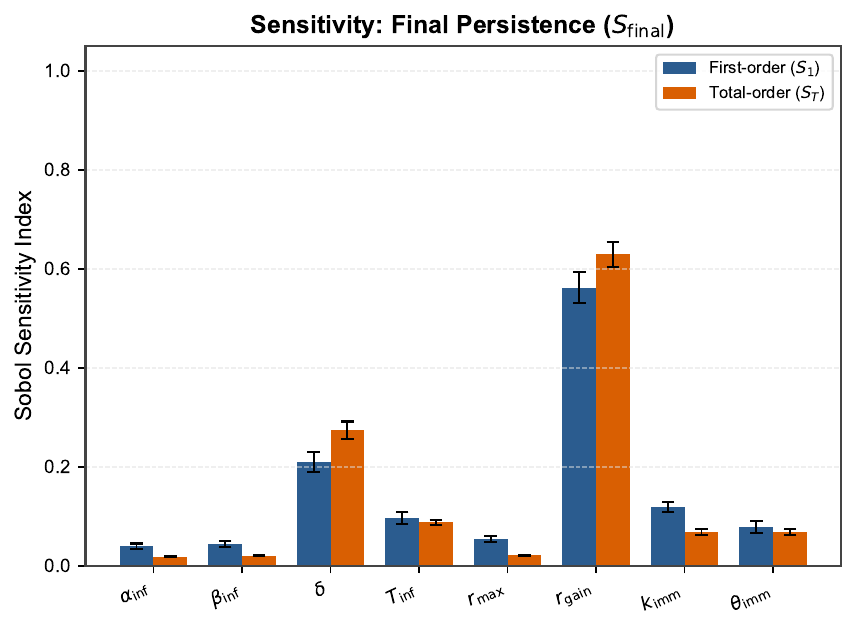}
  \end{subfigure}
  \hfill
  \begin{subfigure}[b]{0.49\textwidth}
    \centering
    \includegraphics[width=\textwidth]{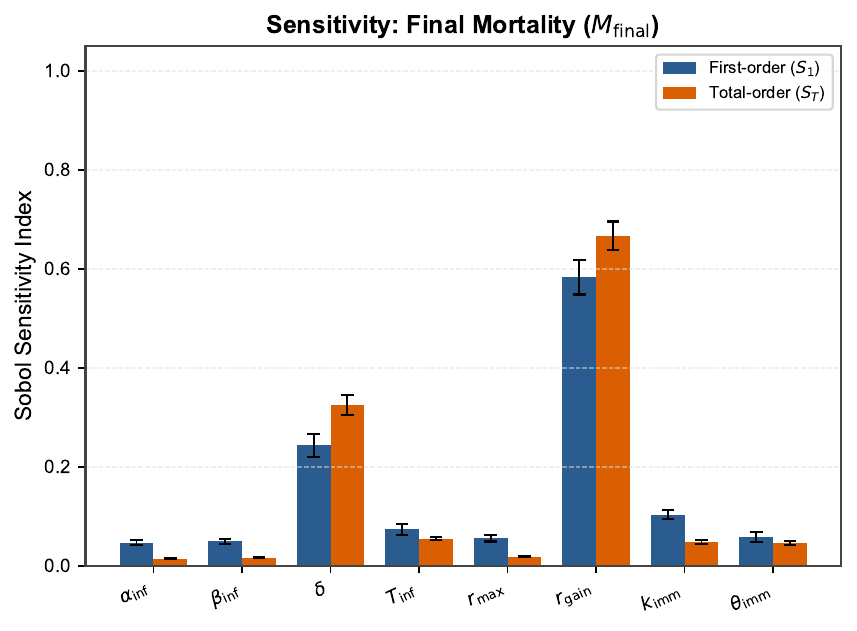}
  \end{subfigure}
  
  \caption{Sobol' analysis results of North Midwest sites.}
  \label{fig:sobolLAT1NMW}
\end{figure}

\begin{figure}[H]
  \centering
  
  \begin{subfigure}[b]{0.49\textwidth}
    \centering
    \includegraphics[width=\textwidth]{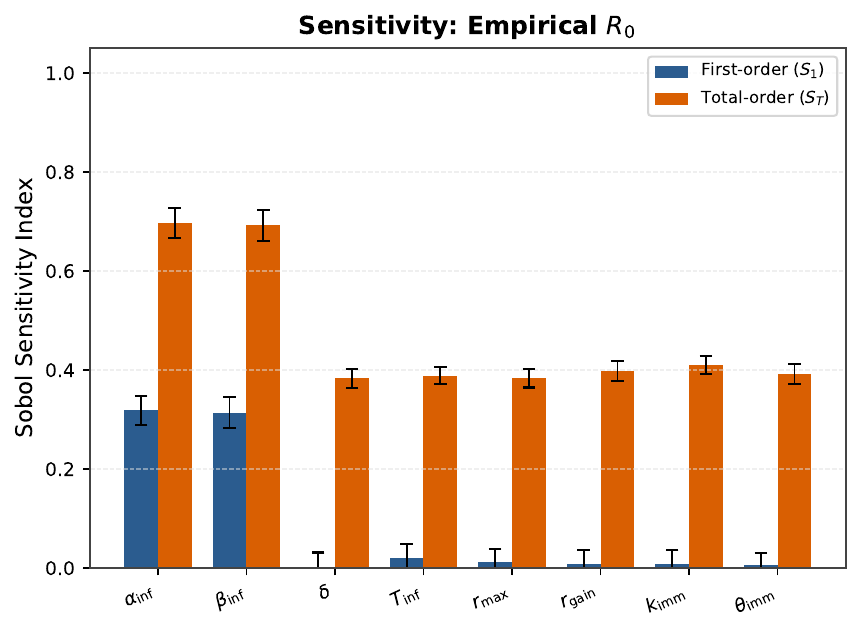}
  \end{subfigure}
  \hfill
  \begin{subfigure}[b]{0.49\textwidth}
    \centering
    \includegraphics[width=\textwidth]{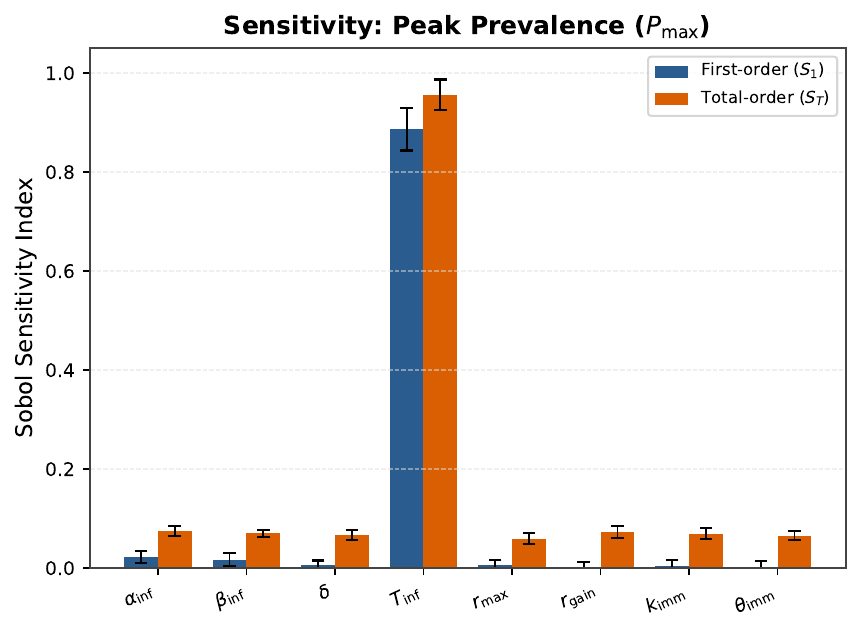}
  \end{subfigure}
  
  \medskip 
  
  \begin{subfigure}[b]{0.49\textwidth}
    \centering
    \includegraphics[width=\textwidth]{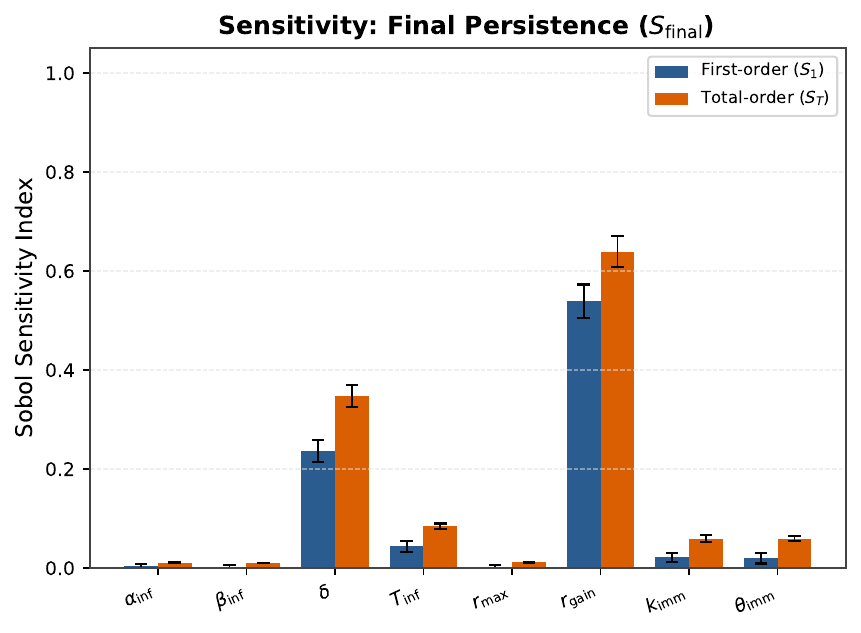}
  \end{subfigure}
  \hfill
  \begin{subfigure}[b]{0.49\textwidth}
    \centering
    \includegraphics[width=\textwidth]{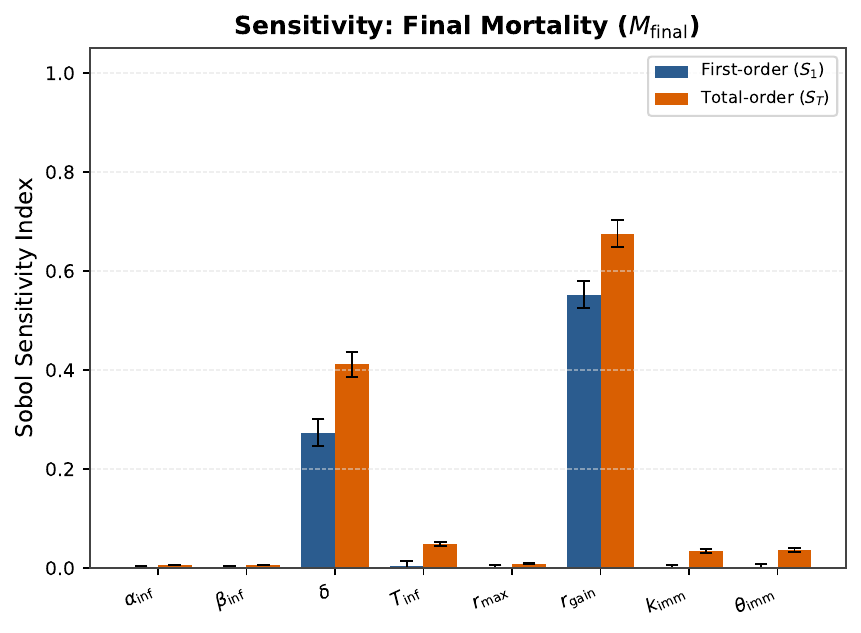}
  \end{subfigure}
  
  \caption{Sobol' analysis results of South Midwest sites.}
  \label{fig:sobolLAT2NMW}
\end{figure}

Figures \ref{fig:sobolLAT1NMW} and
\ref{fig:sobolLAT2NMW} summarize the resulting
sensitivity profiles for the North and South
Midwest, respectively. Each figure presents
first-order and total-order sensitivity indices
for the four output metrics: $R_0$,
$P_{\mathrm{max}}$, $S_{\mathrm{final}}$,
and $M_{\mathrm{final}}$.

The sensitivity patterns indicate that different
parameters influence different aspects of disease
dynamics. The transmission parameters
$\alpha_{\mathrm{inf}}$ and $\beta_{\mathrm{inf}}$
are particularly important for $R_0$, whereas
infection duration ($T_{\mathrm{inf}}$)
plays a major role in peak prevalence.
Long-term population outcomes show a different
sensitivity profile, with acquired resistance
($r_{\mathrm{gain}}$) being the most influential
parameter, followed by environmental spore decay
($\delta$). In comparison, the parameters governing
temporary immunity and inherited resistance
generally exhibit lower sensitivity.
The North and South Midwest sites show broadly
similar parameter rankings across all four metrics.
However, environmental spore decay ($\delta$) has a greater influence on final population persistence and final cumulative mortality in South Midwest sites,
particularly in its total-order sensitivity indices.

\subsection{Invasion Scenarios}

To examine how the initial proportion of infected bats affects epidemic trajectories, we simulated multiple ``invasion scenarios''. 
In these experiments, we varied the initial infection proportions ($1\%$, $5\%$, $10\%$, and $25\%$) while keeping the initial
population size fixed at $100$ bats.
These invasion trajectories were tested under two environmental decay regimes ($\delta=0.01$ and $\delta=0.05$, corresponding to daily environmental fungal decay rates of 1\% and 5\%, respectively).
We then tracked simulations across a period of 20 years ($T=365 \times 20$) to observe whether small initial introductions still lead to collapse, or whether the environmental ``cleaning'' capacity of the hibernaculum against \textit{P. destructans} buffers the population from a drastic collapse. We plot a running average over 20 days to smooth short-term stochastic fluctuations.

\begin{figure}[H]
  \centering
  \includegraphics[width=0.49\linewidth]{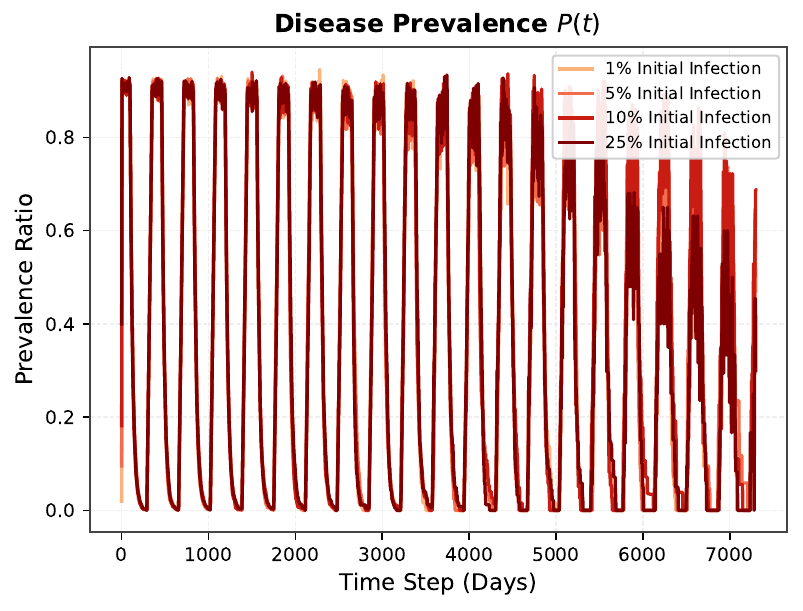}\hfill
  \includegraphics[width=0.49\linewidth]{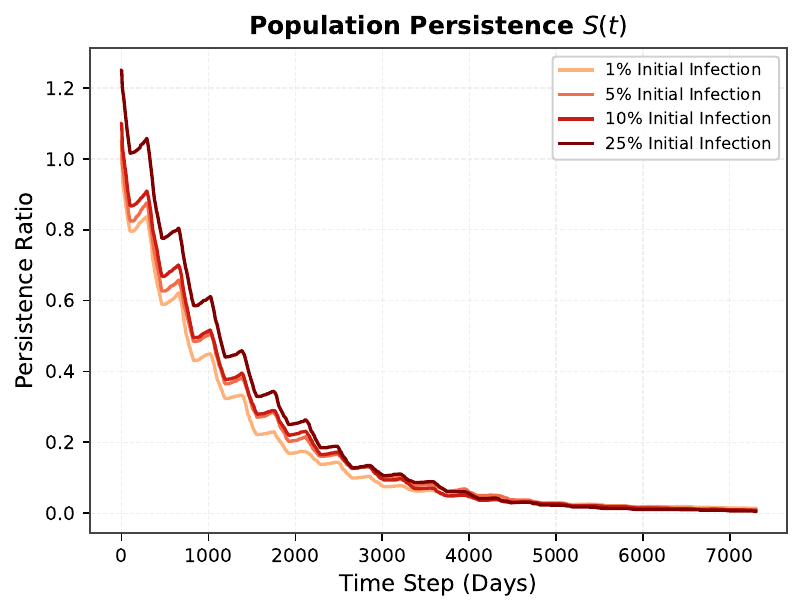}
  \caption{Invasion scenarios for the North Midwest with environmental fungal decay rate $\delta=0.01$.}
  \label{fig:smalldelta:LAT1NMW}
\end{figure}

The North and South Midwest clusters exhibited
similar invasion trajectories under both
environmental decay regimes, with nearly identical results for the lower fungal decay rate
($\delta=0.01$). Therefore, we present only
the North Midwest results for this decay rate
(Figure \ref{fig:smalldelta:LAT1NMW}).
The trajectories show a similar rate of population
decline across initial infection proportions,
with population persistence decreasing by approximately half every 1,000 days over the displayed time period.

\begin{figure}[H]
  \centering
  \includegraphics[width=0.49\linewidth]{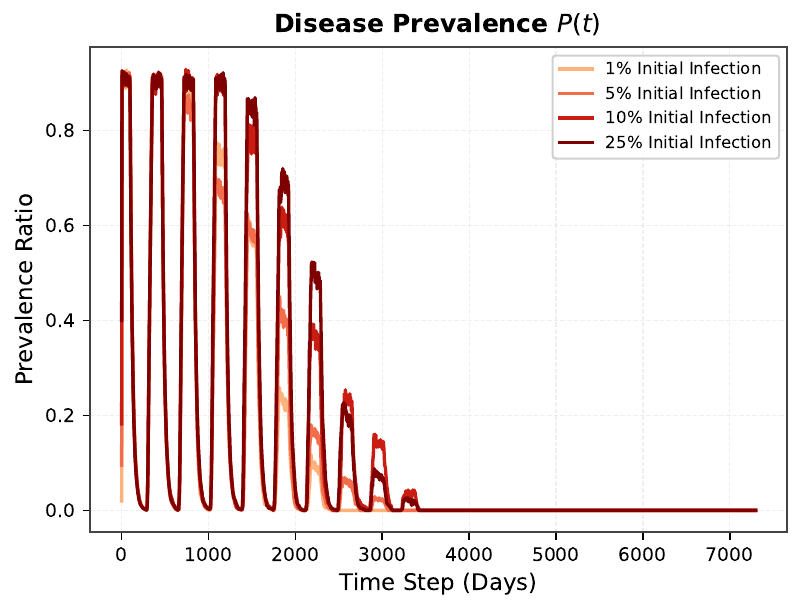}\hfill
  \includegraphics[width=0.49\linewidth]{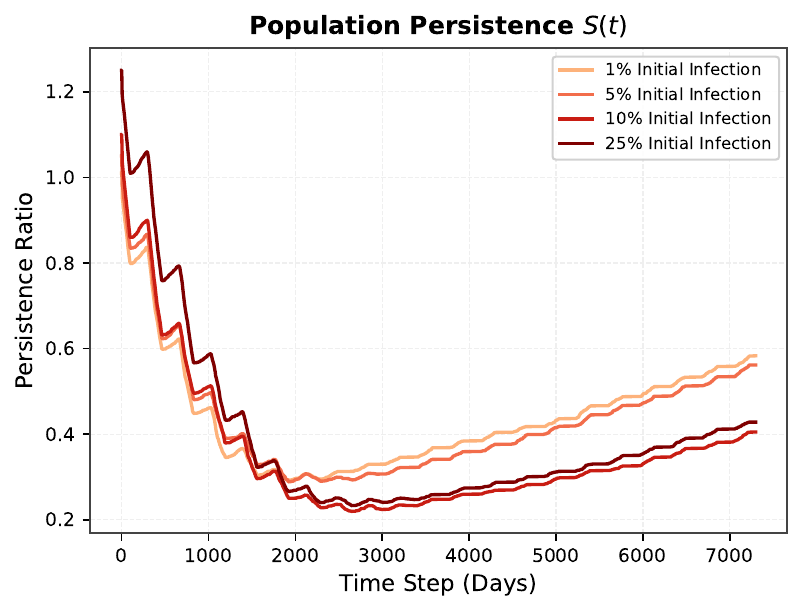}
  \caption{Invasion scenarios for the North Midwest with environmental fungal decay rate $\delta=0.05$.}
  \label{fig:largedelta:LAT1NMW}
\end{figure}

\begin{figure}[H]
  \centering
  \includegraphics[width=0.49\linewidth]{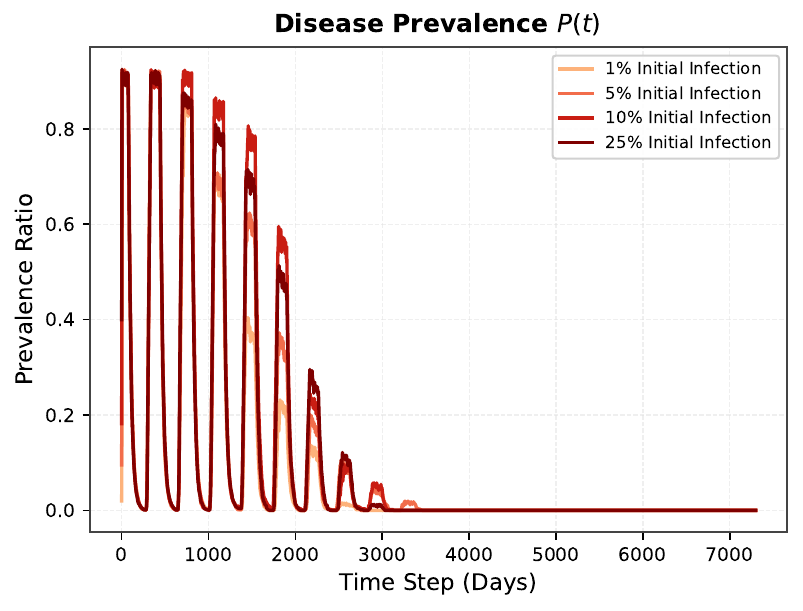}\hfill
  \includegraphics[width=0.49\linewidth]{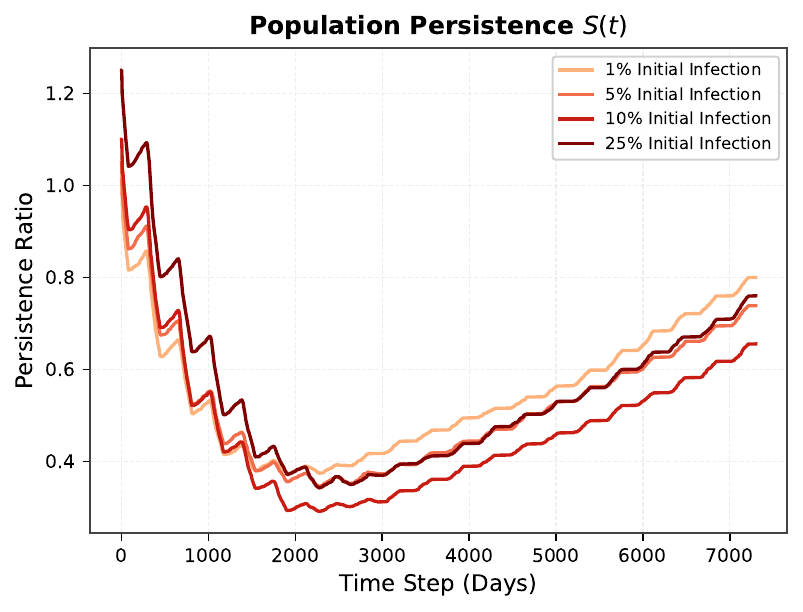}
  \caption{Invasion scenarios for the South Midwest with environmental fungal decay rate $\delta=0.05$.}
  \label{fig:largedelta:LAT2SMW}
\end{figure}

However, for the larger fungal decay rate
($\delta=0.05$), the two regional clusters
exhibit some differences. Both populations
reach minimum persistence around day $2,000$.
The minimum persistence values range from
approximately $0.2$ to $0.3$ for North Midwest
sites (Figure \ref{fig:largedelta:LAT1NMW})
and from approximately $0.3$ to $0.4$ for South
Midwest sites (Figure \ref{fig:largedelta:LAT2SMW}).
Both regional clusters subsequently exhibit
population recovery, with the South Midwest
sites reaching higher final persistence values.


\subsection{Parameter Phase Diagrams}

To map the global behavior of the system across continuous
parameter space, we constructed phase diagrams examining
the effects of simultaneously changing two chosen parameters.
By sweeping across pairs of dominant parameters identified
in the Sobol' analysis, we tracked population trajectories
at different times. We chose two pairs of parameters:
(1) $\delta$ vs. $r_{\mathrm{gain}}$, as both are
prominent drivers of final persistence and mortality,
examined at years $1$, $3$, $5$, and $10$; and
(2) $T_{\mathrm{inf}}$ vs. $\delta$, to examine how
parameters associated with peak infection and long-term
population outcomes jointly influence population dynamics,
examined at years $1$, $3$, $10$, and $20$. Representative snapshots of the chosen years are shown below.

For each parameter pair, we generated 3D surfaces and
2D heatmaps to visualize population outcomes.
We found the simulation results to be noisy, even on a $20 \times 20$ parameter grid, so each
``year'' represents the average of the output over
all 365 days of that year. Additionally, each pair
of parameter values in the grid was simulated across
four independent stochastic realizations, and the
results at each grid cell were averaged over all four
runs. Finally, we applied slight Gaussian smoothing
for visual clarity ($\sigma=0.3$ for both the 3D surfaces
and 2D heatmaps).

The North and South Midwest phase diagrams exhibited
similar qualitative patterns for both parameter pairs.
Therefore, we present only the North Midwest results
to avoid redundancy. The diagrams illustrate how
population outcomes vary across parameter combinations
and time scales. 

\begin{figure}[H]
  \centering
  
  \begin{subfigure}[b]{0.49\textwidth}
    \centering
    \includegraphics[width=\textwidth]{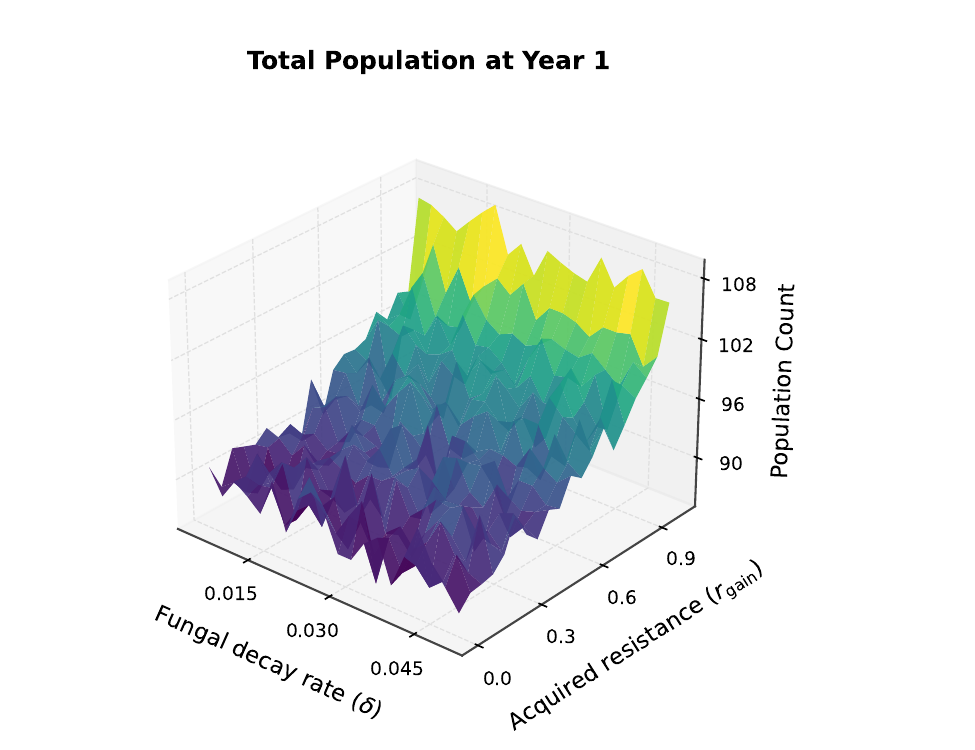}
  \end{subfigure}
  \hfill
  \begin{subfigure}[b]{0.49\textwidth}
    \centering
    \includegraphics[width=\textwidth]{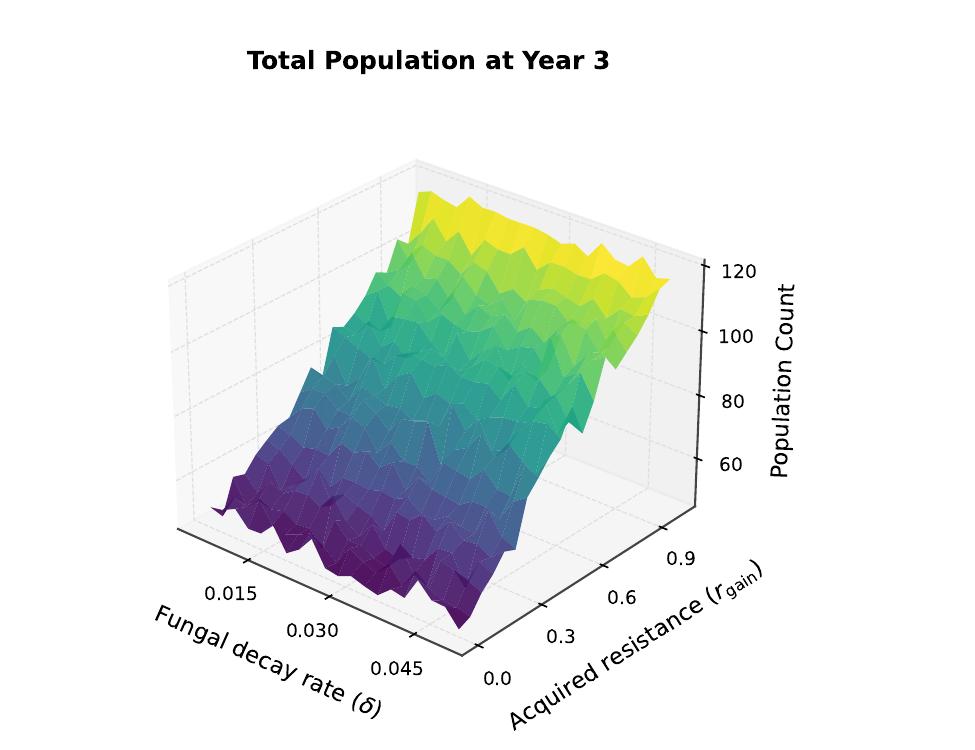}
  \end{subfigure}
  
  \medskip 
  
  \begin{subfigure}[b]{0.49\textwidth}
    \centering
    \includegraphics[width=\textwidth]{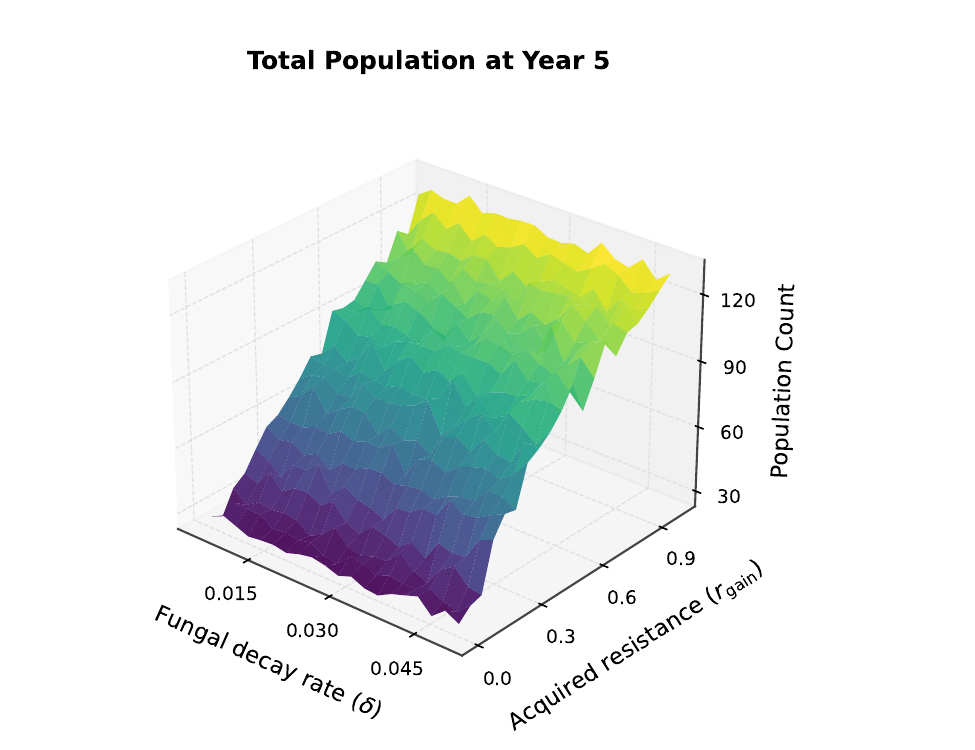}
  \end{subfigure}
  \hfill
  \begin{subfigure}[b]{0.49\textwidth}
    \centering
    \includegraphics[width=\textwidth]{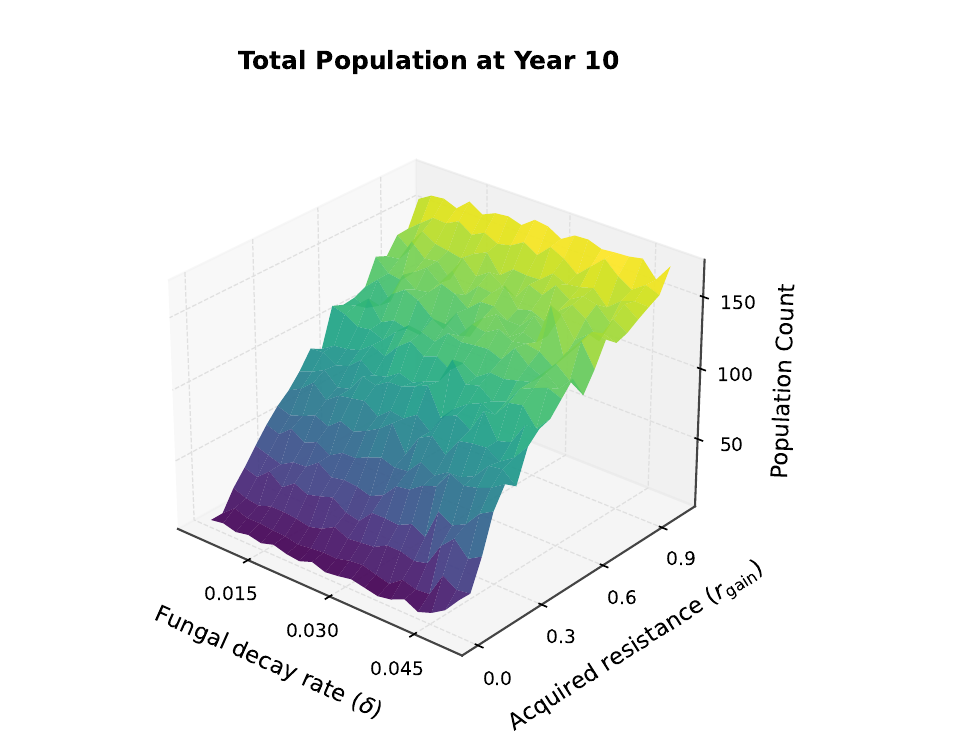}
  \end{subfigure}
\caption{North Midwest phase diagrams showing the
effects of environmental fungal decay rate
($\delta$) and resistance gain ($r_{\mathrm{gain}}$)
on total bat population size at years 1, 3, 5,
and 10.}

  \label{fig:phasediagram_finalpersistenceLAT1NMW}
\end{figure}

\begin{figure}[H]
    \centering
    \includegraphics[width=\linewidth]{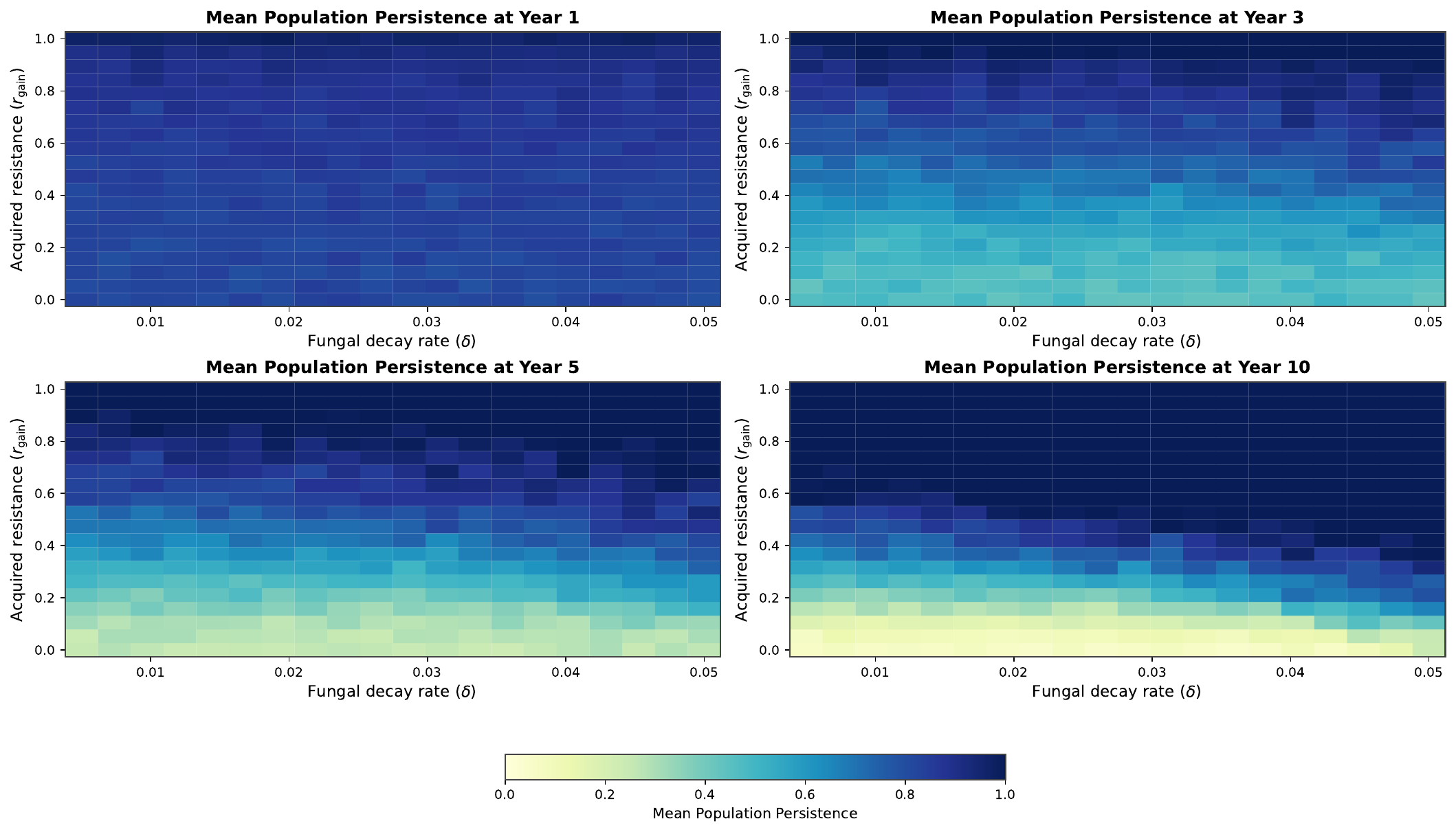}
\caption{North Midwest heatmaps showing the effects of
environmental fungal decay rate ($\delta$) and
resistance gain ($r_{\mathrm{gain}}$) on bat
population persistence at years 1, 3, 5, and 10.}

    \label{fig:heatmap_finalpersistenceLAT1NMW}
\end{figure}

The phase diagrams comparing the fungal decay rate $\delta$ and acquired resistance parameter $r_{\mathrm{gain}}$ (Figure \ref{fig:phasediagram_finalpersistenceLAT1NMW}) show a broadly consistent relationship between these two parameters influencing population persistence across the selected years, where $\delta$ has a minor effect on population size compared with $r_{\mathrm{gain}}$. The corresponding heat map (Figure \ref{fig:heatmap_finalpersistenceLAT1NMW}) shows similar results, with considerable stochastic
variability in year 1 and a clearer positive effect
of acquired resistance on population persistence
in years 3 and 5. By year 10, higher fungal decay
rates also contribute to greater population
persistence. \\

\begin{figure}[H]
  \centering  
  \begin{subfigure}[b]{0.49\textwidth}
    \centering
    \includegraphics[width=\textwidth]{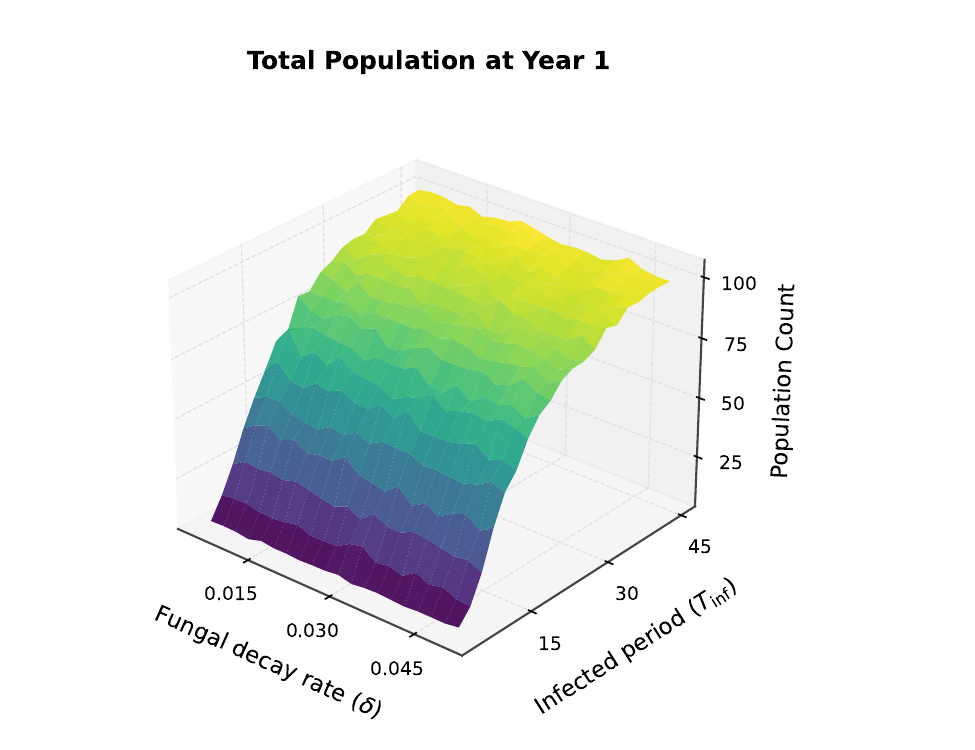}
  \end{subfigure}
  \hfill
  \begin{subfigure}[b]{0.49\textwidth}
    \centering
    \includegraphics[width=\textwidth]{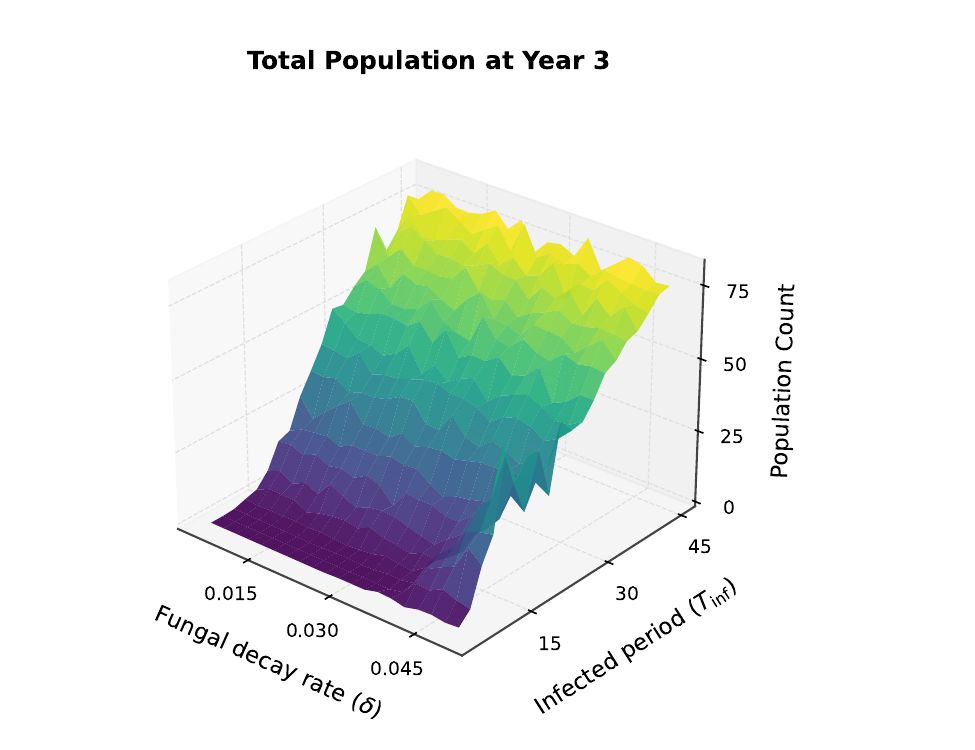}
  \end{subfigure}
  
  \medskip 
  
  \begin{subfigure}[b]{0.49\textwidth}
    \centering
    \includegraphics[width=\textwidth]{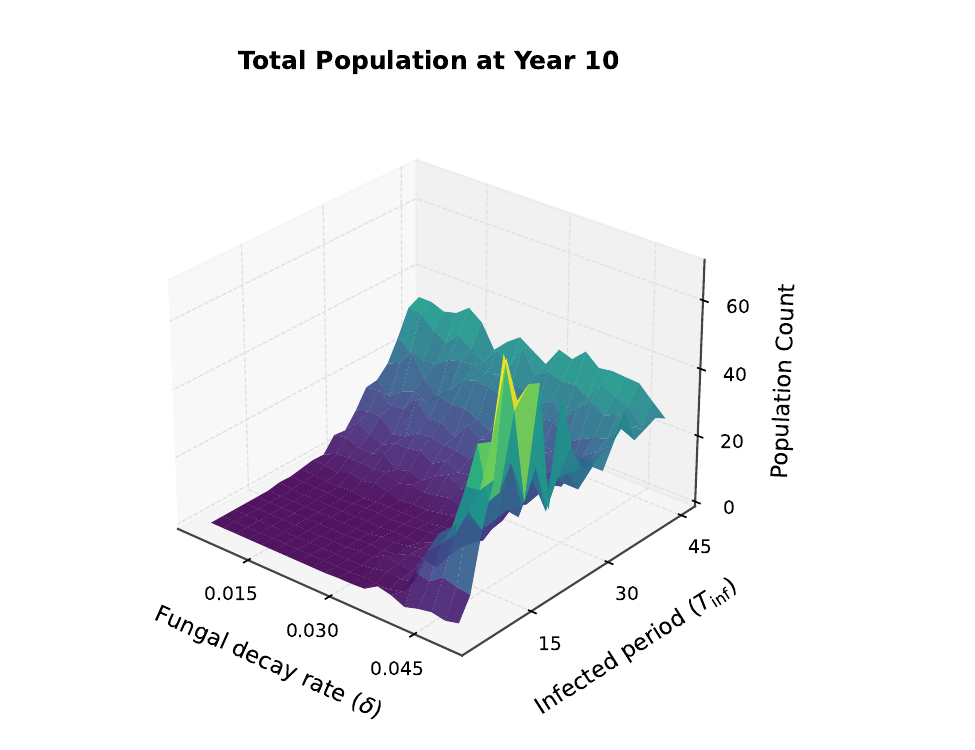}
  \end{subfigure}
  \hfill
  \begin{subfigure}[b]{0.49\textwidth}
    \centering
    \includegraphics[width=\textwidth]{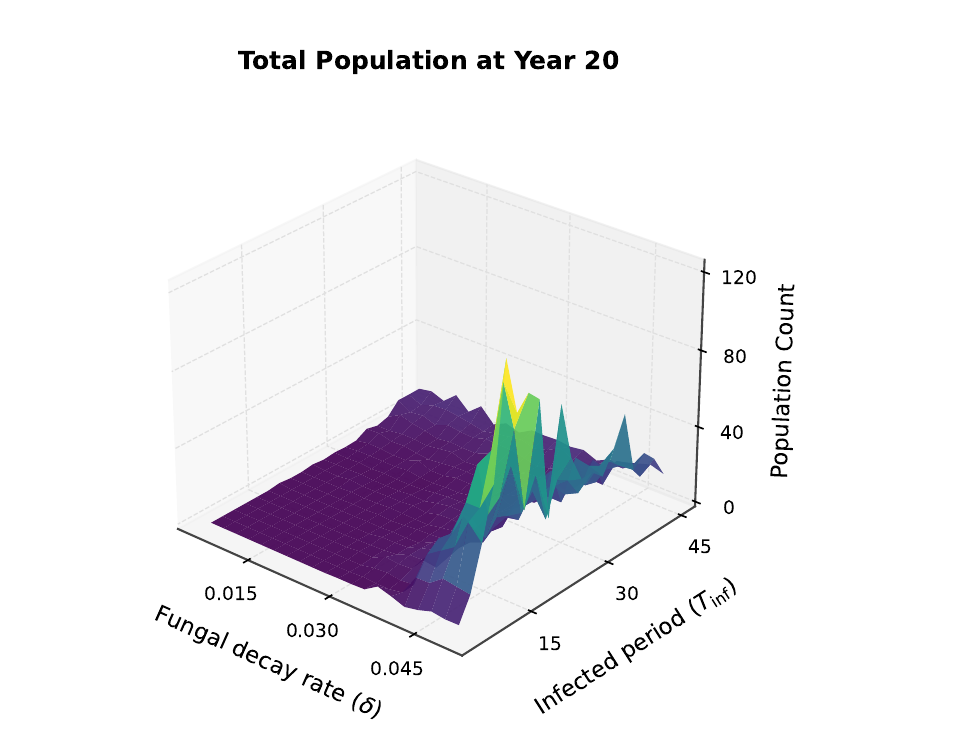}
  \end{subfigure}
\caption{North Midwest phase diagrams showing the
effects of infection duration ($T_{\mathrm{inf}}$)
and environmental fungal decay rate ($\delta$)
on total bat population size at years 1, 3,
10, and 20.}

  \label{fig:phasediagram_longevityLAT1NMW}
\end{figure}

\begin{figure}[H]
    \centering
    \includegraphics[width=\linewidth]{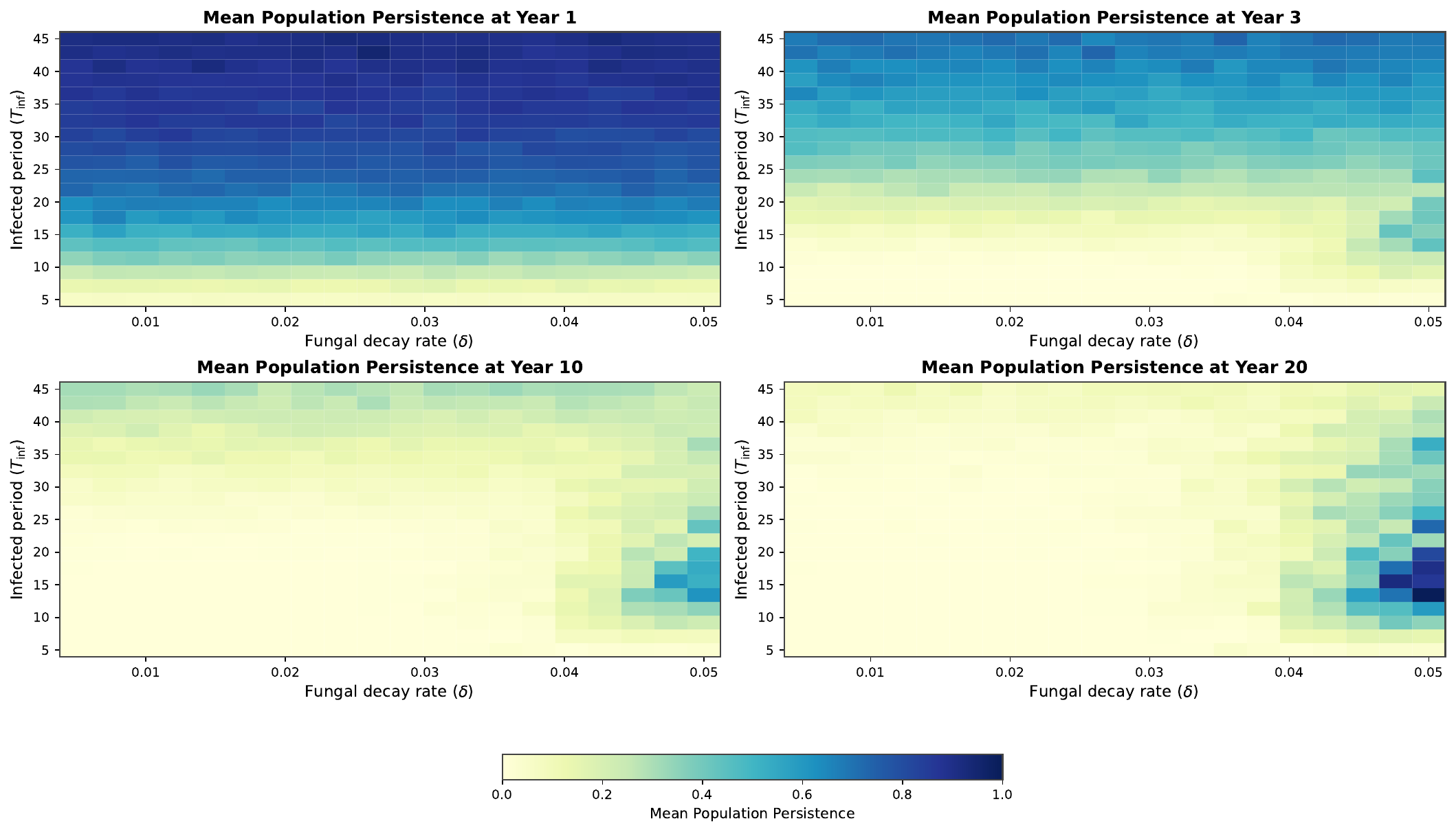}
\caption{North Midwest heatmaps showing the effects of
infection duration ($T_{\mathrm{inf}}$) and
environmental fungal decay rate ($\delta$) on
bat population persistence at years 1, 3, 10, and 20.}

  \label{fig:heatmap_longevityLAT1NMW}
\end{figure}

The phase diagrams for infection duration $T_{\mathrm{inf}}$ and
fungal decay rate $\delta$ (Figure \ref{fig:phasediagram_longevityLAT1NMW}) show population declines over time
for certain parameter combinations, particularly
at low environmental fungal decay rates ($\delta$). 
These declines are visible by year 3 but become
substantially more pronounced by years 10 and 20, as seen in the corresponding heat map (Figure \ref{fig:heatmap_longevityLAT1NMW}).

\section{Discussion}\label{sec:discussion}

The importance of environmental fungal persistence
for long-term population outcomes in our simulations
is consistent with previous findings on WNS.
In \cite{hoyt_environmental_2020} it was demonstrated
that environmental pathogen loads and their persistence
play an important role in infection dynamics and
population declines. Similarly,
\cite{langwig_invasion_2015} documented the accumulation
of environmental contamination following pathogen
invasion in Midwest hibernacula, with earlier infections
and more severe population declines during the
subsequent winter. Thus, the sensitivity of long-term
population persistence and cumulative mortality to
environmental fungal decay ($\delta$) is biologically plausible and
consistent with the established role of
environmental conditions in sustaining transmission.

The phase diagrams and invasion scenarios both highlight the importance of considering longer simulation periods when evaluating population outcomes. In particular, the temporal phase diagrams for \(T_{\mathrm{inf}}\) and \(\delta\) (Figure \ref{fig:phasediagram_longevityLAT1NMW}) suggest that population trajectories may appear relatively similar during the first several years, even when their longer-term outcomes differ substantially. For certain parameter combinations, pronounced declines become apparent only after a decade or more. This delayed response bears a conceptual resemblance to \textit{extinction debt}, in which population declines or extinctions occur long after the environmental changes that initiate them \cite{tilman_habitat_1994,kuussaari_extinction_2009}. Although our single-hibernaculum simulations do not establish extinction debt in its conventional multi-patch metapopulational sense, they illustrate a related challenge for conservation: short-term observations may conceal longer-term population vulnerability. At the same time, the delay between initial infection and substantial population decline may provide an opportunity for conservation interventions before more severe losses occur.

Although acquired resistance ($r_{\mathrm{gain}}$)
is the most influential parameter, environmental
fungal decay ($\delta$) also plays an important
role, particularly in the South Midwest, whereas
inherited resistance and temporary immunity
parameters show comparatively little influence.

Host immunological mechanisms in WNS remain poorly characterized, especially whether surviving bats gain lasting resistance, experience only a brief post-recovery protection, pass down resistance to offspring, or some combination of these. Mathematical modeling provides a controlled way to test the effects of each to determine which hypothetical immunological mechanism has the most impact on populations, without requiring expensive field studies. When acquired resistance ($r_{\mathrm{gain}}$)
is weak or absent, environmental fungal decay
($\delta$) becomes a major determinant of
long-term population persistence. The comparatively small influence of inherited
resistance and temporary immunity in our model
suggests that acquired resistance warrants
particular attention in future immunological studies.

The invasion scenarios suggest that the relationship
between initial infection conditions and long-term
population outcomes depends on environmental fungal
decay. Regardless of low ($\delta=0.01$) or high ($\delta=0.05$) decay rates, populations decline for the first several years, at approximately the same rate regardless of how many initial bats were infected. At low decay rates ($\delta=0.01$),
population declines are similar across the initial
infection proportions examined, with total extinction by years 15-20 (Figure \ref{fig:smalldelta:LAT1NMW}). At higher decay
rates ($\delta=0.05$), population trajectories
show slightly greater variation, with no clear trend based on how many initial bats were infected in North Midwest sites (Figure \ref{fig:largedelta:LAT1NMW}), and lower rates of initial infection performing slightly better in South Midwest sites (Figure \ref{fig:largedelta:LAT2SMW}). At the higher fungal decay rate ($\delta=0.05$),
populations reach minimum persistence around year 5
and subsequently recover. These findings suggest
that environmental conditions may influence
the consequences of initial disease introduction, while the proportion of initial bats infected has little to no influence on long-term population dynamics.

\section{Conclusions}\label{sec:conclusions}

In this study, we developed a BN model to investigate the effects of WNS on bat populations in Midwestern hibernacula. The model incorporates infection, recovery, mortality, immunity, and environmental fungal contamination, allowing us to examine population dynamics under different epidemiological and environmental conditions. We calibrated population-related parameters using observed hibernaculum data and investigated the model's behavior through global sensitivity analysis, invasion scenarios, and parameter phase diagrams.

Our sensitivity analysis identified different parameters influencing short-term disease dynamics and long-term population outcomes. Transmission parameters were most influential for the empirical reproduction number, while infection duration had the greatest influence on peak prevalence. For long-term population persistence and cumulative mortality, acquired resistance was the most influential parameter. For the same metrics, environmental fungal decay was the second most influential parameter. These findings highlight regional differences in parameter sensitivity and the importance of both host-acquired resistance and environmental fungal persistence in the long-term dynamics of WNS within our model.

The invasion scenarios and parameter phase diagrams provide additional perspectives on how initial infection levels, environmental decay, and immunity may influence population trajectories over time. The invasion scenarios indicate that the effects of an initial outbreak depend heavily on environmental fungal decay, with the size of the initial outbreak having minimal impact on long-term survivability. The phase diagrams further show that population declines can become substantially more pronounced over extended simulation periods, emphasizing the importance of considering long-term population outcomes.

Overall, our results demonstrate the usefulness of the BN framework for exploring interactions among infection processes, host responses, and environmental conditions in WNS-affected bat populations. The model provides a basis for investigating which mechanisms most strongly influence population outcomes and may help inform future studies of disease persistence and bat population conservation. Specifically, we emphasize the need for short-term colony observations to be paired with multi-year tracking to identify delayed population declines associated with WNS. We also highlight which of the poorly understood host traits matter most at the colony level, showing that when bats do not develop lasting acquired resistance, environmental fungal decay becomes a primary determinant of long-term population survival within our model, guiding the immunological and environmental measurements that may be most beneficial to collect.

Future research could extend the BN
model by incorporating additional environmental
characteristics of hibernacula, particularly
temperature, humidity, and their effects on
fungal persistence and disease transmission.
It could also incorporate broader climatic changes,
such as drought and shorter winters.
Further investigation of the interactions
among transmission intensity, acquired resistance,
and mortality, including the time required for bats
to develop immunity following infection, could help
clarify the mechanisms underlying long-term
population persistence.
Additional regional data and comparisons across
hibernacula could also help explain differences
in population responses and improve the model's
applicability to diverse environmental conditions.
The framework could also be extended to other bat
species by accounting for differences in contact
rates and clustering behavior during hibernation.
These extensions could provide further insight
into WNS dynamics and inform the development
of conservation strategies.


\section*{Acknowledgments}

We thank Dr. Jeremy White (University of Nebraska at Omaha) for his ecological
expertise and guidance during the initial development of the model.

This work was completed utilizing the Holland Computing Center of the University
of Nebraska, which receives support from the University of Nebraska at Lincoln Office of Research and Innovation, and the Nebraska Research Initiative.

During preparation of this manuscript, the authors unexpectedly acquired an additional bat observation when a wild bat apparently established residence on the last author's balcony. The observation was not included in the analysis.

\section*{Declarations}

\subsection*{Data Availability Statement}
The datasets generated or analyzed during the current study
are publicly available in the first author's GitHub repository at
\href{https://github.com/calvm32/Boolean_Networks_Ecology}
{https://github.com/calvm32/Boolean\_Networks\_Ecology}.

\subsection*{Conflict of Interest Statement}
The authors declare that they have no conflicts of interest.

\subsection*{Author Contribution Statement}

C.~H. contributed to ecological literature synthesis, 
mathematical model formulation, reviewed model alignment to ecological processes, designed and implemented the final 
computational model and parallel simulation pipeline, 
and performed all formal simulation experiments and numerical analyses.
N.~L. contributed to ecological literature synthesis, mathematical model formulation, reviewed model alignment to ecological processes, performed ecological data collection and curation, and early exploratory computational prototyping.
M.T.~V. contributed to the Boolean network formulation
and description, literature-based selection of
probability distributions and parameter ranges,
and research supervision.
All authors contributed to manuscript writing,
reviewing, and revision, and approved the final manuscript.

\subsection*{Tool and Computational Resource Disclosure}

Artificial intelligence (AI) tools (Claude, Anthropic;
Gemini, Google; ChatGPT, OpenAI) were used to assist with some code development, literature searches, manuscript 
editing, and revision. The authors independently reviewed
and verified the resulting codes, mathematical claims,
and references and take full responsibility for the
content of the manuscript.


\newpage
\bibliographystyle{unsrt}
\bibliography{references}\label{refs}

\end{document}